\documentclass[aps,prc,letterpaper,11pt,twoside,tightenlines,nofootinbib,showpacs,preprint]{revtex4-1}
\usepackage{epsfig}
\usepackage{subfigure}
\usepackage{amsmath}
\begin{document}
\arraycolsep1.5pt
\newcommand{\Ima}{\textrm{Im}}
\newcommand{\Rea}{\textrm{Re}}
\newcommand{\mev}{\textrm{ MeV}}
\newcommand{\gev}{\textrm{ GeV}}

\def\bra#1#2#3#4#5{_{(\ell_M={#4}/2,\,\ell_B={#5})}\big\langle S_{c\bar
    c}={#1},\, {\cal L}=\frac{{#2}}{2}\,;  J=\frac{{#3}}{2}|}
\def\bracc#1#2#3#4{_{(\ell_M=0,\,\ell_B=\frac{{#4}}{2})}\big\langle S_{c\bar
    c}={#1},\, {\cal L}=\frac{{#2}}{2}\,; J=\frac{{#3}}{2}|}
\def\ket#1#2#3#4#5{| S_{c\bar c}={#1},\, {\cal L}=\frac{{#2}}{2}\,;J=\frac{{#3}}{2}\big\rangle_{(\ell_M={#4}/2,\,\ell_B={#5})}}
\def\ketcc#1#2#3#4{| S_{c\bar c}={#1},\, {\cal L}=\frac{{#2}}{2}\,; J=\frac{{#3}}{2}\big\rangle_{(\ell_M=0,\,\ell_B=\frac{{#4}}{2})}}

\title{Combining heavy quark spin and local hidden gauge symmetries in the dynamical generation of hidden charm baryons}

\author{C. W. Xiao$^{1,2}$, J. Nieves$^{2}$ and E. Oset$^{1,2}$}
\affiliation{
$^{1}$Departamento de F\'{\i}sica Te\'orica, Universidad de Valencia, Spain\\
$^{2}$IFIC, Centro Mixto Universidad de 
Valencia-CSIC,
Institutos de Investigaci\'on de Paterna, Aptdo. 22085, 46071 Valencia,
Spain
}

\date{\today}

\begin{abstract}

We present a coupled channel unitary approach to obtain states dynamically generated from the meson baryon interaction with hidden charm, using constraints of heavy quark spin symmetry. We use as basis of states, $\bar D B$, $\bar D^* B$ states, with $B$ baryon charmed states belonging to the 20 representations of SU(4) with $J^P=1/2^+,~3/2^+$. In addition we also include the $\eta_c N$ and $J/\psi N$ states. The inclusion of these coupled channels is demanded by heavy quark spin symmetry, since in the large $m_Q$ limit the $D$ and $D^*$ states are degenerate and are obtained from each other by means of a spin rotation, under which QCD is invariant. The novelty in the work is that we use dynamics from the extrapolation of the local hidden gauge model to SU(4) and we show that this dynamics fully respects the constraints of heavy quark spin symmetry. 

With the full space of states demanded by the heavy quark spin symmetry and the dynamics of the local hidden gauge we look for states dynamically generated and find four basic states which are bound, corresponding to $\bar D \Sigma_c$, $\bar D \Sigma_c^*$, $\bar D^* \Sigma_c$  and $\bar D^* \Sigma_c^*$, decaying mostly into $\eta_c N$ and $J/\psi N$. All the states appear in isospin $I=1/2$ and we find no bound states or resonances in $I=3/2$. The $\bar D \Sigma_c$ state appears in $J=1/2$, the $\bar D \Sigma_c^*$ in $J=3/2$, the  $\bar D^* \Sigma_c$ appears nearly degenerate in $J=1/2, ~3/2$ and the $\bar D^* \Sigma_c^*$ appears nearly degenerate in $J=1/2, ~3/2, ~5/2$, with the peculiarity that in $J=5/2$ the state has zero width in the space of states chosen. All the states are bound with about 50 MeV with respect to the corresponding $\bar D B$ thresholds and the width, except for the $J=5/2$ state, is also of the same order of magnitude. 
 
\end{abstract}

\maketitle

\section{Introduction}
\label{Intro} 
In this paper we investigate hidden charm baryons which come from the interaction of mesons with baryons, with the system containing a $c \bar c$ component. This can come from pseudoscalar-baryon or vector-baryon interactions. In \cite{wuprl,wuprc} this problem was faced and, mostly by means of the $\bar D \Sigma_c$, $\bar D \Lambda_c$ and $\bar D^* \Sigma_c$, $\bar D^* \Lambda_c$ components, a series of meson-baryon dynamically generated, relatively narrow $N^*$ and $\Lambda^*$ resonances, were predicted around 4.3 GeV. The interaction used in \cite{wuprl,wuprc} was obtained from an extrapolation to SU(4), conveniently broken, of the local hidden gauge dynamics used for SU(3) \cite{hidden1,hidden2,hidden4}. 

The local hidden gauge model is dynamically very rich and is considered a good representation of QCD at low energies. In the pseudoscalar sector it contains the lowest order chiral Lagrangian \cite{Weinberg:1978kz,Gasser:1983yg} and, in addition, the hidden gauge Lagrangian provides the interaction between vectors and their coupling to pseudoscalars. It implements the vector meson dominance hypothesis of Sakurai \cite{sakurai} and, within this assumption, it also provides the second order Lagrangian for pseudoscalar-pseudoscalar interaction of \cite{Gasser:1983yg}, as shown in \cite{Ecker:1989yg}. The use of the local hidden gauge Lagrangian in connection with coupled channels and unitary techniques provides a tool that allows to study vector meson interactions in the intermediate energy range where the interaction itself gives rise to dynamically generated states. This is the case for the $\rho \rho$ interaction, from where one obtains the $f_2(1270)$ and $f_0(1370)$ resonances \cite{raquelrho} and its extension to the interactions of vectors of the  $\rho$ nonet \cite{gengvec}, from where a few more dynamically generated resonances are obtained, like the $f_0(1710)$, $f'_2(1525)$ and $K^*_2(1430)$. The properties of the resonances obtained are shown to be consistent with the radiative decay to two photons \cite{junko} and to two-photon and one photon-one vector meson in \cite{tanyageng}. Similarly, consistency with experiment has been shown in $J/\psi \to  \phi(\omega) R$ \cite{conchinos}, with R any of the resonances of \cite{gengvec}, and in $J/\psi$ radiative decays in \cite{radiative}. The extension of these ideas to the charm and hidden charm sector have also shown that some of the excited $D$ states and X,Y,Z states recently reported could be explained in terms of molecules involving mesons with charm \cite{raquelhideko,xyz,raqueltanya,Dong:2012hc,Branz:2010gd,Branz:2010sh,Lee:2009hy}.

  The extension of the local hidden gauge approach to the baryon sector for the interaction of vector mesons with baryons has also been tackled: the interaction of vector mesons with the decuplet of baryons is studied in \cite{sourav} and with the octet of baryons in \cite{angelsvec}. In both cases some dynamically generated resonances are obtained which can be associated to reported resonances in the PDG \cite{pdg}. One step forward in this direction is the consideration of vector-baryon and pseudoscalar-baryon simultaneously in the interaction, which has been done in \cite{javier}. A thorough work in this direction has also been done in  \cite{kanchan1,kanchan2,kanchan3}. A review of the hidden gauge approach for vector-baryon and vector-nucleus interaction can be seen in \cite{review}. 
  
  Work in the charm sector for meson-baryon interaction has been done along different lines, which share similarities with the local hidden gauge approach \cite{Lutz:2003jw,Lutz:2005ip,Hofmann:2005sw,Hofmann:2006qx}. A different approach is done in \cite{laura}, where one uses the analogy of the work of the $\bar K N$ interaction and replaces a s-quark by the c-quark. As mentioned in \cite{Mizutani:2006vq}, while the potentials obtained are fine with this prescription, in the coupled channel approach one is missing channels that mix charm and strangeness in that approach. In \cite{Mizutani:2006vq,JimenezTejero:2009vq} the work of \cite{Lutz:2003jw} is retaken and appropriate modifications are done in the potentials and the regularization scheme. Similar work is also done by the J\"ulich group in \cite{Haidenbauer:2007jq,Haidenbauer:2008ff,Haidenbauer:2010ch}. All these works share the dynamical generation of the $\Lambda_c(2595)$, which comes mostly from the interaction of the $D N$ channel. Some hidden charm baryonic state is also generated in  \cite{Hofmann:2005sw}, albeit with a binding of the order of 1000 MeV, difficult to accommodate with the generated potentials as discussed in  \cite{wuprl,wuprc}. 

As we can see, the topic of baryonic molecules with charm and hidden charm has attracted much attention, and the coming of the FAIR facility is certainly stimulating much work along these lines. Yet, an element missing in principle in these works is the consideration of heavy quark spin symmetry, which should be a good symmetry when working with mesons and baryons with charm. From the point of view of Heavy Quark Spin Symmetry (HQSS), which is a proper QCD  spin-flavor symmetry~\cite{IW89,Ne94,MW00} when the quark masses become much larger than the typical confinement scale, $\Lambda_{\rm QCD}$, one should consider in the same footing $D$ and $D^*$ as well as charmed members of the $20$ SU(4) representation of baryons containing the octet of the proton and the $20$ representation containing the decuplet of the $\Delta$ when their isospin and strange contents are the same. Work along these lines is done in \cite{GarciaRecio:2008dp,Gamermann:2010zz,Romanets:2012hm,Garcia-Recio:2013gaa}. In these references, an extended Weinberg-Tomozawa (WT) interaction to four flavors is derived. The model for four flavors includes all basic hadrons (pseudoscalar and vector mesons, and $\frac12^+$ and $\frac32^+$ baryons) and it reduces to the WT interaction in the sector where Goldstone bosons are involved, while it incorporates HQSS in the sector where charm quarks participate. Charmed and strange baryons are studied in \cite{Romanets:2012hm}, where among other results, a heavy-quark spin symmetry doublet is associated to the tree stars $\Xi_c(2790)$ and $\Xi_c(2815)$ pair of resonances. Moreover, the model derived in Ref.~\cite{Romanets:2012hm} also accommodates naturally the three stars charmed resonances $\Lambda_c(2595)$ and $\Lambda_c (2625)$. The $\Lambda_c (2595)$ was previously dynamically generated in other schemes based on $t$-channel vector-meson-exchange models \cite{Hofmann:2005sw,laura,Mizutani:2006vq,JimenezTejero:2009vq}, but in \cite{Romanets:2012hm}, as first pointed out in \cite{GarciaRecio:2008dp}, a large (dominant) $ND^*$ component in its structure was claimed. This is in sharp contrast with the findings of the former references, where it was generated mostly as one $ND$ bound state, since the $ND^*$ channel was not considered in the coupled channels space. The work of Ref.~\cite{GarciaRecio:2012db} takes advantage of the underlying spin-flavor extended WT structure of the couplings of the model of Refs. \cite{GarciaRecio:2008dp,Romanets:2012hm} and it is used to study odd parity bottom-flavored baryon resonances by replacing a $c-$quark by a $b-$quark.\footnote{The universality of the interactions of heavy quarks, regardless of their concrete (large) mass, flavor and spin state, follows from QCD \cite{IW89,Ne94,MW00}.} Two resonances $\Lambda_b(5912)$ and $\Lambda_b(5920)$, which are heavy quark spin symmetry partners, are predicted in \cite{GarciaRecio:2012db} and turn out to be in excellent agreement with the two narrow baryon resonances with beauty recently observed by the LHCb Collaboration~\cite{Aaij:2012da}. Finally, in \cite{Garcia-Recio:2013gaa} the model of Ref.~\cite{Romanets:2012hm} is extended to the hidden charm sector, subject of the current work. Seven odd parity $N-$like and three $\Delta-$like states with masses around 4 GeV, most of them as bound states, are predicted in \cite{Garcia-Recio:2013gaa}. These states form heavy-quark spin multiplets, which are almost degenerate in mass. As we will discuss below, here we will extend the hidden gauge approach, and the predictions found in this work will notably differ from those obtained in \cite{Romanets:2012hm} (we do not obtain any isospin 3/2 states, and the isospin 1/2 states are significantly heavier, in the region of 4.4 GeV). Besides the use of different dynamics, both consistent as we shall see with the leading order HQSS requirements, the scheme to renormalize the Bethe-Salpeter equation employed here is also quite different to that advocated in \cite{Garcia-Recio:2013gaa}, which for the case of the hidden charm sector leads to appreciable differences. We will give some more details when our results will be presented.

However, the HQSS does not determine the potential, simply puts some constraints in it, so the determination in the works of \cite{GarciaRecio:2008dp,Gamermann:2010zz,Romanets:2012hm,Garcia-Recio:2013gaa,GarciaRecio:2012db} is made assuming extra elements of SU(8) spin-isospin symmetry. The work with baryons along these lines has run parallel to work in the meson sector \cite{Nieves:2012tt,HidalgoDuque:2012pq,HidalgoDuque:2012ej,Guo:2009id,Guo:2013sya}. In these works, an Effective Field Theory (EFT) that implements leading order (LO) HQSS constrains is constructed and its consequences are derived. Many dynamically generated resonances are obtained as HQSS partners of the $X(3872)$, $Z_b(10610)$, and the $Z_b(10650)$, some of which can be associated to known resonances, but most are predictions. 

   In the present work we come back to the local hidden gauge approach and introduce $D^*$ and the members of the $20$-plet of the $\Delta$, as demanded by HQSS, but the dynamics linking the different pseudoscalar-baryon and vector-baryon states is taken from the hidden gauge approach. We look again in the hidden charm baryon sector. What we find in the work is that the matrix elements obtained with the dynamics of the local hidden gauge approach respect the HQSS for the dominant terms in the mass of the heavy quarks, something that was not known so far. Another of the findings is that, within this model, the transition from $D$ to $D^*$ states is subleading in the heavy quark mass counting, as well as the transition from the $1/2^+$ baryons of the $20$ representation to those of the $3/2^+$ $20$ representation. In this sense, the findings of the present work give extra support to earlier works using the local hidden gauge approach where the different spaces were not allowed to connect. Yet, in addition to the states obtained in  \cite{wuprl,wuprc} from $D B_{1/2^+}$ and $D^* B_{1/2^+}$, one obtains extra states from the $D B_{3/2^+}$ and $D^* B_{3/2^+}$, which will be reported here.

\section{Lowest order HQSS constraints}

HQSS predicts that all types of spin interactions vanish for infinitely
massive quarks: the dynamics is unchanged under arbitrary
transformations of the spin of the heavy quark ($Q$).  The
spin-dependent interactions are proportional to the chromomagnetic
moment of the heavy quark, and hence, they are of the order of $1/m_Q$. The
total angular momentum $\vec{J}$ of the hadron is always a conserved
quantity, but in this case the spin of the heavy quark $\vec{S}_Q$ is also
conserved in the $m_Q\to \infty$ limit. Consequently, the spin of the
light degrees of freedom $\vec{S}_l=\vec{J}-\vec{S}_Q$ is a conserved quantity in that limit. Thus, heavy hadrons come in doublets (unless $s_l=0$), 
 containing states with total spin
$j_\pm=s_l\pm 1/2$ (with $\vec{S}_l^2=s_l(s_l+1)$ and $J^2= j (j+1)$) obtained by combining the spin of the light degrees
of freedom with the spin of the heavy quark $s_Q=1/2$. These doublets
are degenerate in the $m_Q\to \infty$ limit. This is the case for the
ground state mesons $D$ and $D^*$ or $D_s$ and $D^*_s$ which are
composed of a charm quark with $s_Q=1/2$ and light degrees of freedom
with $s_l=1/2$, forming a multiplet of negative parity hadrons with 
spin 0 and 1. The entire multiplet of degenerate states should be
treated in any HQSS inspired formalism as a single field that
transforms linearly under the heavy quark symmetries~\cite{Ne94,MW00}.
For finite charm quark mass, the pseudoscalar and vector $D$ meson
masses differ in about just one pion mass (actually one has  $m_D-m_{D^*} = {\cal O}(1/(m_D+m_{D^*}))$), even less for the strange charmed
mesons, thus it is reasonable to expect that the
coupling $DN \to D^*N$ might play an important role. This is indeed what happens
when SU(8) symmetry is used \cite{GarciaRecio:2008dp,Romanets:2012hm}. Conversely, we shall see that with the local hidden gauge dynamics, the transition $DN \to D^*N$, which is mediated by pion exchange, is rather small and vanishes formally in the limit of zero difference between the mass of the  $D$ and the $D^*$.  Something similar occurs with the transition from the $1/2^+$ baryons to those with  $3/2^+$. As a consequence, four diagonal blocks develop when the hidden gauge dynamics is used while at the same time the relations due to heavy quark symmetry are exactly fulfilled in each of the blocks. With a different dynamics than the one provided by the SU(8) symmetry, the numerical results that we obtain are also different from those obtained in \cite{Garcia-Recio:2013gaa} and we will make a discussion of these results in the present work.

We study baryons with hidden charm and $I=1/2,\ 3/2,\ J=1/2,\ 3/2,\ 5/2$. We take as coupled channels states with $\eta_c,\ J/\psi$ and a $N$ or a $\Delta$, and states with $\bar D,\ \bar D^*$ and $\Lambda_c,\ \Sigma_c$ or $\Sigma_c^*$. For the different $I,\ J$ quantum numbers we have the following space states. \\

1) $J=1/2,\ I=1/2$

$\quad \eta_c N,\ J/\psi N,\ \bar{D} \Lambda_c,\ \bar{D} \Sigma_c,\ \bar{D}^* \Lambda_c,\ \bar{D}^* \Sigma_c,\ \bar{D}^* \Sigma_c^*$. \\

2) $J=1/2,\ I=3/2$

$\quad J/\psi \Delta,\ \bar{D} \Sigma_c,\ \bar{D}^* \Sigma_c,\ \bar{D}^* \Sigma_c^*$. \\

3) $J=3/2,\ I=1/2$

$\quad J/\psi N,\ \bar{D}^* \Lambda_c,\ \bar{D}^* \Sigma_c,\ \bar{D} \Sigma_c^*,\ \bar{D}^* \Sigma_c^*$. \\

4) $J=3/2,\ I=3/2$

$\quad \eta_c \Delta,\ J/\psi \Delta,\ \bar{D}^* \Sigma_c,\ \bar{D} \Sigma_c^*,\ \bar{D}^* \Sigma_c^*$. \\

5) $J=5/2,\ I=1/2$

$\quad \bar{D}^* \Sigma_c^*$. \\

6) $J=5/2,\ I=3/2$

$\quad J/\psi \Delta,\ \bar{D}^* \Sigma_c^*$. \\

Attending to the spin quantum number we have thus 17 orthogonal states in the physical basis. Next we will introduce a different basis, that we will call HQSS basis, for which it is straightforward to implement the LO HQSS constraints. In the HQSS basis we will classify the states in terms of the quantum numbers, $J$: total spin of the meson-baryon system, ${\cal L}$: total spin of the light quarks system, $S_{c \bar{c}}$: total spin of the  $c \bar{c}$ subsystem, $\ell_M$: total spin of the light quarks in the meson and $\ell_B$: total spin of the light quarks in the baryon. Note that we assume that all orbital angular momenta are zero, since we are dealing with ground state baryons.

Thus, the 17 orthogonal states in the HQSS basis are given by
\begin{itemize}
\item {\small $\ketcc 0111$, $\ket 01110$, \\ $\ket 01111$}
\item {\small $\ketcc 1111$, $\ket 11110$, \\ $\ket 11111$}
\item {\small $\ketcc 1131$, $\ket 11310$, \\ $\ket 11311$}
\item {\small $\ketcc 0333$,  $\ket 03311$}
\item {\small $\ketcc 1313$, $\ket 13111$}
\item {\small $\ketcc 1333$, $\ket 13311$}
\item {\small $\ketcc 1353$, $\ket 13511$}
\end{itemize}
The approximate HQSS of QCD leads (neglecting ${\cal
  O}(\Lambda_{QCD}/m_Q)$ corrections) to important simplifications when
the HQSS basis is used:
\begin{equation}
\begin{split}
& _{(\ell_M',\ell_B')}\big\langle S'_{c\bar c},\, {\cal L}'; J',\, \alpha'|H^{QCD}|
S_{c\bar c},\, {\cal L}; J,\, \alpha \big \rangle_{(\ell_M,\ell_B)} \\
= \, & \delta_{\alpha \alpha'}\delta_{JJ'}\delta_{S'_{c\bar c}S_{c\bar c} }
\delta_{{\cal L}{\cal L}'}  \big\langle \ell_M'\ell_B' {\cal L};
\alpha ||H^{QCD}  || \ell_M\ell_B {\cal L}; \alpha \big\rangle \ , \label{eq:hqs}
\end{split}
\end{equation}
where $\alpha$ stands for other quantum numbers (isospin and
hypercharge), which are conserved by QCD. Note that the reduced matrix elements do not depend on $S_{c\bar c}$, because QCD dynamics is invariant under separate spin rotations of the charm quark and antiquark. Thus, one can transform a $c \bar{c}$ spin singlet state into a spin triplet state by means of a rotation that commutes with $H^{QCD}$, i.e. a zero cost of energy. Thus, in a given $\alpha$ sector, we have a total of nine unknown low energy constants (LEC's):
\begin{itemize}
\item Three LEC's associated to ${\cal L}=3/2$
\begin{eqnarray}
\lambda_1^\alpha &=& \big\langle \ell_M'=0,\, \ell_B'=\frac32,\, {\cal L}=3/2;\alpha
 ||H^{QCD}  || \ell_M=0,\, \ell_B=\frac32,\, {\cal L}=3/2;\alpha
 \big\rangle \\
\lambda_2^\alpha &=& \big\langle \ell_M'=1/2,\, \ell_B'=1,\, {\cal L}=3/2;\alpha
 ||H^{QCD}  || \ell_M=1/2,\, \ell_B=1,\, {\cal L}=3/2;\alpha
 \big\rangle \\
\lambda_{12}^\alpha &=& \big\langle \ell_M'=0,\, \ell_B'=\frac32,\, {\cal L}=3/2;\alpha
 ||H^{QCD}  ||\ell_M=1/2,\, \ell_B=1,\, {\cal L}=3/2;\alpha
 \big\rangle
\end{eqnarray}
\item Six LEC's associated to ${\cal L}=1/2$
\begin{eqnarray}
\mu_1^\alpha &=& \big\langle \ell_M'=0,\, \ell_B'=\frac12,\, {\cal L}=1/2;\alpha
 ||H^{QCD}  || \ell_M=0,\, \ell_B=\frac12,\, {\cal L}=1/2;\alpha
 \big\rangle \\
\mu_2^\alpha &=& \big\langle \ell_M'=1/2,\, \ell_B'=0,\, {\cal L}=1/2;\alpha
 ||H^{QCD}  || \ell_M=1/2,\, \ell_B=0,\, {\cal L}=1/2;\alpha
 \big\rangle\\
\mu_3^\alpha &=& \big\langle \ell_M'=1/2,\, \ell_B'=1,\, {\cal L}=1/2;\alpha
 ||H^{QCD}  || \ell_M=1/2,\, \ell_B=1,\, {\cal L}=1/2;\alpha
 \big\rangle\\
\mu_{12}^\alpha &=& \big\langle \ell_M'=0,\, \ell_B'=\frac12,\, {\cal L}=1/2;\alpha
 ||H^{QCD}  ||\ell_M=1/2,\, \ell_B=0,\, {\cal L}=1/2;\alpha
 \big\rangle\\
\mu_{13}^\alpha &=& \big\langle \ell_M'=0,\, \ell_B'=\frac12,\, {\cal L}=1/2;\alpha
 ||H^{QCD}  ||\ell_M=1/2,\, \ell_B=1,\, {\cal L}=1/2;\alpha
 \big\rangle\\
\mu_{23}^\alpha &=& \big\langle \ell_M'=1/2,\, \ell_B'=0,\, {\cal L}=1/2;\alpha
 ||H^{QCD}  || \ell_M=1/2,\, \ell_B=1,\, {\cal L}=1/2;\alpha \big\rangle
\end{eqnarray}
\end{itemize}
This means that in the HQSS basis, the  $H^{QCD}$ is a block diagonal matrix,
i.e, up to ${\cal  O}(\Lambda_{QCD}/m_Q)$ corrections, $H^{QCD}=
{\rm Diag}(\mu^\alpha,\mu^\alpha,\mu^\alpha,\lambda^\alpha,\lambda^\alpha,\lambda^\alpha,\lambda^\alpha)$,
where $\mu^\alpha$ and $\lambda^\alpha$ are symmetric matrices of
dimension 3 and 2, respectively.

To exploit Eq.~(\ref{eq:hqs}), one should express hidden charm
uncoupled meson--baryon states in terms of the HQSS basis.
For those states composed of hidden charm mesons ($\ell_M=0$) the
relations  are trivial,
\begin{eqnarray}
|\eta_c N; J=\frac12\big\rangle &=& \ketcc 0111  \\
\nonumber\\ 
|\eta_c \Delta; J=\frac32\big\rangle &=& \ketcc 0333  \\
\nonumber\\ 
|J_\Psi N; J=\frac12\big\rangle &=& \ketcc 1111  \\
\nonumber\\ 
|J_\Psi N; J=\frac32\big\rangle &=& \ketcc 1131  \\
\nonumber\\ 
|J_\Psi \Delta; J=\frac12\big\rangle &=& \ketcc 1313  \\
\nonumber\\ 
|J_\Psi \Delta; J=\frac32\big\rangle &=& \ketcc 1333  \\
\nonumber\\ 
|J_\Psi \Delta; J=\frac52\big\rangle &=& \ketcc 1353
\end{eqnarray}
while for the other states, one needs to use 9-j symbols.

The 9-j symbols are used to relate two basis where the angular momentums are coupled in a different way. Taking two particles with $\vec{l}_1,\ \vec{s}_1$ and $\vec{l}_2,\ \vec{s}_2$, we can combine them to $\vec{j}_1,\ \vec{j}_2$ and finally $\vec{j}_1,\ \vec{j}_2$ to total $\vec{J}$. Alternatively we can couple $\vec{l}_1,\ \vec{l}_2$ to $\vec{L}$, $\vec{s}_1,\ \vec{s}_2$ to $\vec{S}$, and then $\vec{L},\ \vec{S}$ to total $\vec{J}$. These two bases are related as \cite{Rose}
\begin{equation}
\begin{split}
|l_1 s_1 j_1; l_2 s_2 j_2; J M \big\rangle =& \sum_{S,L} [ (2 S + 1) (2 L + 1) (2 j_1 + 1) (2 j_2 + 1)]^{1/2} \\
& \times \left\{
\begin{array}{ccc}
l_1 & l_2 & L \\
s_1 & s_2 & S \\
j_1 & j_2 & J
\end{array}
\right\} \ |l_1 l_2 L; s_1 s_2 S; J M \big\rangle,
\end{split}
\end{equation}
where the symbol $\{ \}$ stands for the 9-j coefficients.

\begin{figure}[tb]
\epsfig{file=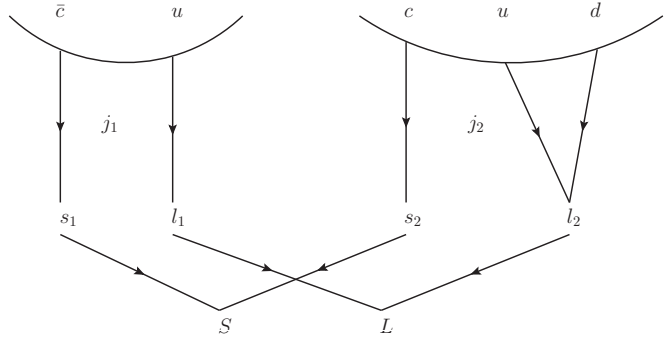, width=9cm}
\caption{Diagrams for the 9-j coefficients evaluation.}%
\label{fig:9j}%
\end{figure}

As an example take a meson(M)-baryon(B) state of the type $\bar{D}^{(*)} B_c$ and look at the recombination scheme on Fig. \ref{fig:9j}.
Thus in this case we have the correspondence,
\begin{align*}
&\text{generic:} && l_1 && l_2 && s_1 && s_2 && j_1 && j_2 && L && S && J  \\
&\text{HQSS:} && \ell_M(\frac{1}{2}) && \ell_B && \frac{1}{2} && \frac{1}{2} && J_M(0,1) && J_B (\frac{1}{2},\frac{3}{2}) && {\cal L} && S_{c \bar{c}} && J (\frac{1}{2},\frac{3}{2},\frac{5}{2}) \ .
\end{align*}
with $J_M$ and $J_B$ the total spin of the meson and baryon respectively. Then one easily finds:
\begin{itemize}
\item $J=1/2$
\begin{eqnarray}
|\bar D \Lambda_c\big \rangle &=& \frac12 \ket01110 \nonumber \\
&+& \frac{\sqrt{3}}{2} \ket11110 \\
\nonumber \\
|\bar D \Sigma_c\big \rangle &=& \frac12 \ket01111 -
\frac{1}{2\sqrt{3}} \ket11111 \nonumber \\ 
&+& \sqrt{\frac23} \ket13111 \\
\nonumber \\
|\bar D^* \Lambda_c\big \rangle &=& \frac{\sqrt{3}}2 \ket01110 \nonumber \\
&-& \frac12 \ket11110\\
\nonumber \\
|\bar D^* \Sigma_c\big \rangle &=& -\frac1{\sqrt{12}} \ket01111 +
\frac56 \ket11111 \nonumber \\ 
&+& \frac{\sqrt{2}}3 \ket13111\\
\nonumber \\
|\bar D^* \Sigma_c^*\big \rangle &=& \frac2{\sqrt{6}} \ket01111 +
\frac{\sqrt{2}}3  \ket11111 \nonumber \\ 
&-& \frac13 \ket13111
\end{eqnarray}

\item $J=3/2$
\begin{eqnarray}
|\bar D^* \Lambda_c\big \rangle &=&  \ket11310\\
\nonumber \\
|\bar D^* \Sigma_c\big \rangle &=& -\frac1{\sqrt{3}} \ket03311 +
\frac13  \ket11311 \nonumber \\ 
&+& \frac{\sqrt{5}}3 \ket13311 \\
\nonumber \\
|\bar D \Sigma_c^*\big \rangle &=& \frac12   \ket03311  
-\frac1{\sqrt{3}} \ket11311 \nonumber \\ 
&+& \sqrt{\frac5{12}} \ket13311\\
\nonumber \\
|\bar D^* \Sigma_c^*\big \rangle &=& \sqrt{\frac5{12}} \ket03311 +
\frac{\sqrt{5}}3   \ket11311 \nonumber \\ 
&+& \frac16 \ket13311
\end{eqnarray}

\item $J=5/2$
\begin{eqnarray}
|\bar D^* \Sigma_c^*\big \rangle &=& \ket13511
\end{eqnarray}

\end{itemize}
Ignoring hidden strange channels, we find the following interactions
for each sector (these are the most general interactions
  compatible with HQSS):
\newpage  
\begin{itemize}
\item $J=1/2$, $I=1/2$
\[
\left. \phantom{(}
\begin{array}{ccccccc}
\phantom{ \sqrt{\frac{2}{3}} \text{$\mu_{13}$}} & \phantom{\frac{\sqrt{2} \text{$\mu_{13}$}}{3}} & \phantom{\sqrt{\frac{2}{3}} \text{$\mu_{23}$}} & \phantom{\frac{1}{3} \sqrt{\frac{2}{3}} (\text{$\mu_3$}-\text{$\lambda_2 $})} &
   \phantom{\frac{\sqrt{2} \text{$\mu_{23}$}}{3}} & \phantom{\frac{1}{9} \sqrt{2}
   (\text{$\mu_3$}-\text{$\lambda_2 $})} & \phantom{\frac{1}{9}
   (\text{$\lambda_2 $}+8 \text{$\mu_3$})}\\
\eta_c N & J_\Psi N &  \bar D \Lambda_c &  \bar D \Sigma_c &  \bar D^* \Lambda_c
  &  \bar D^* \Sigma_c &  \bar D^* \Sigma^*_c 
\end{array}
\right. \phantom{)_{I=1/2}}
\]
\begin{equation}
\left(
\begin{array}{ccccccc}
 \text{$\mu_1$} & 0 & \frac{\text{$\mu_{12}$}}{2} &
 \frac{\text{$\mu_{13}$}}{2} & \frac{\sqrt{3} \text{$\mu_{12}$}}{2} &
 -\frac{\text{$\mu_{13}$}}{2 \sqrt{3}} & \sqrt{\frac{2}{3}}
 \text{$\mu_{13}$} \\ \\
 0 & \text{$\mu_1$} & \frac{\sqrt{3} \text{$\mu_{12}$}}{2} &
 -\frac{\text{$\mu_{13}$}}{2 \sqrt{3}} & -\frac{\text{$\mu_{12}$}}{2}
 & \frac{5 \text{$\mu_{13}$}}{6} & \frac{\sqrt{2}
   \text{$\mu_{13}$}}{3} \\ \\
 \frac{\text{$\mu_{12}$}}{2} & \frac{\sqrt{3} \text{$\mu_{12}$}}{2} &
 \text{$\mu_2$} & 0 & 0 & \frac{\text{$\mu_{23}$}}{\sqrt{3}} &
 \sqrt{\frac{2}{3}} \text{$\mu_{23}$} \\ \\
 \frac{\text{$\mu_{13}$}}{2} & -\frac{\text{$\mu_{13}$}}{2 \sqrt{3}} & 0 & \frac{1}{3} (2 \text{$\lambda_2 $}+\text{$\mu_3$}) & \frac{\text{$\mu_{23}$}}{\sqrt{3}} & \frac{2 (\text{$\lambda_2$}-\text{$\mu_3$})}{3 \sqrt{3}} & \frac{1}{3} \sqrt{\frac{2}{3}}
 (\text{$\mu_3$}-\text{$\lambda_2 $}) \\ \\
 \frac{\sqrt{3} \text{$\mu_{12}$}}{2} & -\frac{\text{$\mu_{12}$}}{2} & 0 & \frac{\text{$\mu_{23}$}}{\sqrt{3}} & \text{$\mu_2$} & -\frac{2 \text{$\mu_{23}$}}{3} & \frac{\sqrt{2} \text{$\mu_{23}$}}{3} \\ \\
 -\frac{\text{$\mu_{13}$}}{2 \sqrt{3}} & \frac{5 \text{$\mu_{13}$}}{6} & \frac{\text{$\mu_{23}$}}{\sqrt{3}} & \frac{2 (\text{$\lambda_2 $}-\text{$\mu_3$})}{3 \sqrt{3}} & -\frac{2 \text{$\mu_{23}$}}{3} & \frac{1}{9} (2 \text{$\lambda_2 $}+7 \text{$\mu_3$}) &
 \frac{1}{9} \sqrt{2} (\text{$\mu_3$}-\text{$\lambda_2 $}) \\ \\
 \sqrt{\frac{2}{3}} \text{$\mu_{13}$ } & \frac{\sqrt{2} \text{$\mu_{13}$}}{3}\; & \sqrt{\frac{2}{3}} \text{$\mu_{23}$ } & \frac{1}{3} \sqrt{\frac{2}{3}} (\text{$\mu_3$}-\text{$\lambda_2 $})\; &
   \frac{\sqrt{2} \text{$\mu_{23}$}}{3}\;\; & \frac{1}{9} \sqrt{2}
   (\text{$\mu_3$}-\text{$\lambda_2 $}) & \frac{1}{9}
   (\text{$\lambda_2 $}+8 \text{$\mu_3$}) \\ \\
\end{array}
\right)_{ I=1/2}
\label{eq:ji11}
\end{equation}

\item $J=1/2$, $I=3/2$
\[
\left. \phantom{(}
\begin{array}{cccc}
 \phantom{-\frac{\text{$\lambda_{12}$}}{3}} & \phantom{\frac{1}{3} \sqrt{\frac{2}{3}} (\text{$\mu_3$}-\text{$\lambda_2$})} & \phantom{\frac{1}{9} \sqrt{2} (\text{$\mu_3$}-\text{$\lambda_2$})} & \phantom{\frac{1}{9}
   (\text{$\lambda_2$}+8 \text{$\mu_3$})} \\
J_\Psi \Delta &  \bar D \Sigma_c &  \bar D^* \Sigma_c &  \bar D^* \Sigma^*_c 
\end{array}
\right. \phantom{)_{I=3/2}}
\]
\begin{equation}
\left(
\begin{array}{cccc}
 \text{$\lambda_1$} & \sqrt{\frac{2}{3}} \text{$\lambda_{12}$} &
 \frac{\sqrt{2} \text{$\lambda_{12}$}}{3} &
 -\frac{\text{$\lambda_{12}$}}{3} \\ \\
 \sqrt{\frac{2}{3}} \text{$\lambda_{12}$} & \frac{1}{3} (2 \text{$\lambda_2$}+\text{$\mu_3$}) & \frac{2 (\text{$\lambda_2$}-\text{$\mu_3$})}{3 \sqrt{3}} & \frac{1}{3}
   \sqrt{\frac{2}{3}} (\text{$\mu_3$}-\text{$\lambda_2$}) \\ \\
 \frac{\sqrt{2} \text{$\lambda_{12}$}}{3} & \frac{2 (\text{$\lambda_2$}-\text{$\mu_3$})}{3 \sqrt{3}} & \frac{1}{9} (2 \text{$\lambda_2$}+7 \text{$\mu_3$}) & \frac{1}{9} \sqrt{2}
   (\text{$\mu_3$}-\text{$\lambda_2$})\\ \\
 -\frac{\text{$\lambda_{12}$}}{3} & \frac{1}{3} \sqrt{\frac{2}{3}} (\text{$\mu_3$}-\text{$\lambda_2$})\;\; & \frac{1}{9} \sqrt{2} (\text{$\mu_3$}-\text{$\lambda_2$}) & \frac{1}{9}
   (\text{$\lambda_2$}+8 \text{$\mu_3$}) \\
\end{array}
\right)_{ I=3/2}
\label{eq:ji13}
\end{equation}
\newpage
\item  $J=3/2$, $I=1/2$
\[
\left. \phantom{(}
\begin{array}{ccccc}
\phantom{\frac{\sqrt{5} \text{$\mu_{13}$}}{3}} & \phantom{\frac{\sqrt{5}
   \text{$\mu_{23}$}}{3}} & \phantom{\frac{1}{9} \sqrt{5}
 (\text{$\mu_3$}-\text{$\lambda_2 $})} & \phantom{\frac{1}{3} \sqrt{\frac{5}{3}} 
(\text{$\lambda_2$}-\text{$\mu_3$})} & \phantom{\frac{1}{9} (4 \text{$\lambda_2 $}+5 \text{$\mu_3$})}\\
 J_\Psi N &  \bar D^* \Lambda_c &  \bar D^* \Sigma_c 
&  \bar D \Sigma^*_c  &  \bar D^* \Sigma^*_c 
\end{array}
\right. \phantom{)_{I=1/2}}
\]

\begin{equation}
\left(
\begin{array}{ccccc}
 \text{$\mu_1$} & \text{$\mu_{12}$} & \frac{\text{$\mu_{13}$}}{3} & -\frac{\text{$\mu_{13}$}}{\sqrt{3}} & \frac{\sqrt{5} \text{$\mu_{13}$}}{3} \\\\
 \text{$\mu_{12}$} & \text{$\mu_2$} & \frac{\text{$\mu_{23}$}}{3} & -\frac{\text{$\mu_{23}$}}{\sqrt{3}} & \frac{\sqrt{5} \text{$\mu_{23}$}}{3} \\\\
 \frac{\text{$\mu_{13}$}}{3} & \frac{\text{$\mu_{23}$}}{3} & \frac{1}{9} (8 \text{$\lambda_2 $}+\text{$\mu_3$}) & \frac{\text{$\lambda_2 $}-\text{$\mu_3$}}{3 \sqrt{3}} & \frac{1}{9}
   \sqrt{5} (\text{$\mu_3$}-\text{$\lambda_2 $}) \\\\
 -\frac{\text{$\mu_{13}$}}{\sqrt{3}} & -\frac{\text{$\mu_{23}$}}{\sqrt{3}} & \frac{\text{$\lambda_2 $}-\text{$\mu_3$}}{3 \sqrt{3}} & \frac{1}{3} (2 \text{$\lambda_2 $}+\text{$\mu_3$}) &
   \frac{1}{3} \sqrt{\frac{5}{3}} (\text{$\lambda_2 $}-\text{$\mu_3$}) \\\\
 \frac{\sqrt{5} \text{$\mu_{13}$}}{3}\; & \frac{\sqrt{5}
   \text{$\mu_{23}$}}{3}\; & \frac{1}{9} \sqrt{5}
 (\text{$\mu_3$}-\text{$\lambda_2 $})\; & \frac{1}{3} \sqrt{\frac{5}{3}} 
(\text{$\lambda_2$}-\text{$\mu_3$})\; & \frac{1}{9} (4 \text{$\lambda_2 $}+5 \text{$\mu_3$}) \\
\end{array}
\right)_{I=1/2}
\label{eq:ji31}
\end{equation}

\item  $J=3/2$, $I=3/2$
\[
\left. \phantom{(}
\begin{array}{ccccc}
\phantom{\frac{1}{2} \sqrt{\frac{5}{3}} \text{$\lambda_{12} $}} & \phantom{\frac{1}{2} \sqrt{\frac{5}{3}} \text{$\lambda_{12} $}} & \phantom{\frac{1}{9} \sqrt{5} (\text{$\mu_3$}-\text{$\lambda_2 $})} & \phantom{\frac{1}{3} \sqrt{\frac{5}{3}}
   (\text{$\lambda_2 $}-\text{$\mu_3$})} & \phantom{\frac{1}{9} (4 \text{$\lambda_2 $}+5 \text{$\mu_3$})} \\
\eta_c \Delta & J_\Psi \Delta &  \bar D^* \Sigma_c 
&  \bar D \Sigma^*_c  &  \bar D^* \Sigma^*_c 
\end{array}
\right. \phantom{)_{I=3/2}}
\]
\begin{equation}
\left(
\begin{array}{ccccc}
 \text{$\lambda_1 $} & 0 & -\frac{\text{$\lambda_{12} $}}{\sqrt{3}} & \frac{\text{$\lambda_{12} $}}{2} & \frac{1}{2} \sqrt{\frac{5}{3}} \text{$\lambda_{12} $} \\\\
 0 & \text{$\lambda_1 $} & \frac{\sqrt{5} \text{$\lambda_{12} $}}{3} & \frac{1}{2} \sqrt{\frac{5}{3}} \text{$\lambda_{12} $} & \frac{\text{$\lambda_{12} $}}{6} \\\\
 -\frac{\text{$\lambda_{12} $}}{\sqrt{3}} & \frac{\sqrt{5} \text{$\lambda_{12} $}}{3} & \frac{1}{9} (8 \text{$\lambda_2 $}+\text{$\mu_3$}) & \frac{\text{$\lambda_2 $}-\text{$\mu_3$}}{3
   \sqrt{3}} & \frac{1}{9} \sqrt{5} (\text{$\mu_3$}-\text{$\lambda_2 $}) \\\\
 \frac{\text{$\lambda_{12} $}}{2} & \frac{1}{2} \sqrt{\frac{5}{3}} \text{$\lambda_{12} $} & \frac{\text{$\lambda_2 $}-\text{$\mu_3$}}{3 \sqrt{3}} & \frac{1}{3} (2 \text{$\lambda_2$}+\text{$\mu_3$}) & \frac{1}{3} \sqrt{\frac{5}{3}} (\text{$\lambda_2 $}-\text{$\mu_3$}) \\\\
 \frac{1}{2} \sqrt{\frac{5}{3}} \text{$\lambda_{12} $} & \frac{\text{$\lambda_{12} $}}{6} & \frac{1}{9} \sqrt{5} (\text{$\mu_3$}-\text{$\lambda_2 $})\;\; & \frac{1}{3} \sqrt{\frac{5}{3}}
   (\text{$\lambda_2 $}-\text{$\mu_3$})\; & \frac{1}{9} (4 \text{$\lambda_2 $}+5 \text{$\mu_3$}) \\
\end{array}
\right)_{I=3/2}
\label{eq:ji33}
\end{equation}

\item  $J=5/2$, $I=1/2$
\[
\left. \phantom{(}
\begin{array}{c}
\phantom{\text{$\lambda_2$}}\\
 \bar D^* \Sigma^*_c 
\end{array}
\right. \phantom{)_{I=1/2}}
\]
\begin{equation}
\left(
\begin{array}{c}
 \text{$\lambda_2$} \\
\end{array}
\right)_{I=1/2}
\label{eq:ji51}
\end{equation}

\item  $J=5/2$, $I=3/2$
\[
\left. \phantom{(}
\begin{array}{cc}
\phantom{\text{$\lambda_{12}$}} &\phantom{\text{$\lambda_{12}$}} \\
 J_\Psi \Delta \quad & \bar D^* \Sigma^*_c 
\end{array}
\right. \phantom{)_{I=3/2}}
\]
\begin{equation}
\left(
\begin{array}{cc}
 \text{$\lambda_1$} & \text{$\lambda_{12}$} \\\\
 \text{ $\lambda_{12}$  } & \text{ $\lambda_2$ } \\
\end{array}
\right)_{I=3/2}
\label{eq:ji53}
\end{equation}

\end{itemize}

We should stress, once more, that $\mu$ and $\lambda$ depend on isospin, and thus  those LEC's corresponding to $I=1/2$ are not the same as those corresponding to $I=3/2$. Though, they can be related using SU(3) flavor symmetry.

There is a total of 7 (6$\mu's$ and $\lambda_2$) independent LEC's for $I=1/2$, while for $I=3/2$, we have 4 (3$\lambda's$ and $\mu_3$) LEC's. Thus, when one neglects open and hidden strange channels, we have a total of 11 LEC's. The extension of the WT model, using SU(8) spin-flavor symmetry \cite{Garcia-Recio:2013gaa}, provides predictions for all these LEC's. Namely,\footnote{We thank L. L. Salcedo.}
\begin{eqnarray}
I=1/2 &\to& \mu_1=0,\quad
\mu_2=\mu_3=1,\quad \mu_{12}=-\mu_{13}=\sqrt{6},\quad
\mu_{23}=-3,\quad \lambda_2=-2; \\    
I=3/2 &\to& \mu_3=-2, \quad \lambda_1=0,\quad
\lambda_{12}=2\sqrt{3},\quad \lambda_2=4,
\end{eqnarray}
up to an overall $\frac{1}{4f^2} (k^0 + k'^0)$ factor, being $k^0$ and $k'^0$ the center mass energies of the incoming and outgoing mesons.
The extension of the local hidden gauge approach to the charm sector provides different values, as we discuss below.

Note that in \cite{Garcia-Recio:2013gaa} (Sec. II.F) the 12 most general operators allowed by SU(3)$\times$HQSS in the hidden charm baryon-meson sector were already given. Moreover, the reduction of these Lagrangians when no strangeness is involved was also discussed. In this latter case, there are 11 independent couplings, which determine the 11 LEC's ($\mu's$ and $\lambda's$) introduced in Eqs. (\ref{eq:ji11}$-$\ref{eq:ji53}).

\section{Brief description of the local hidden gauge formalism}

We summarize the formalism of the hidden gauge interaction for vector mesons which we take from 
\cite{hidden1,hidden2} (see also useful Feynman rules in \cite{hidekoroca}) extended to SU(4). 
The Lagrangian accounting for the interaction of 
vector mesons amongst themselves is given by
\begin{equation}
{\cal L}_{III}=-\frac{1}{4}\langle V_{\mu \nu}V^{\mu\nu}\rangle \ ,
\label{lVV}
\end{equation}
where the $\langle \rangle$ symbol represents the trace in the SU(4) space 
and $V_{\mu\nu}$ is given by 
\begin{equation}
V_{\mu\nu}=\partial_{\mu} V_\nu -\partial_\nu V_\mu -ig[V_\mu,V_\nu]\ ,
\end{equation}
with the coupling of the theory given by $g=\frac{M_V}{2f}$
where $f=93$~MeV is the pion decay constant. The magnitude $V_\mu$ is the SU(4) 
matrix of the vectors of the meson 15-plet + singlet, given by \cite{Gamermann:2008jh}
\begin{equation}
V_\mu=\left(
\begin{array}{cccc}
\frac{\rho^0}{\sqrt{2}}+\frac{\omega}{\sqrt{2}} & \rho^+ & \quad K^{*+}\quad & \quad \bar D^{*0} \\ & & & \\
 \rho^{-} & -\frac{\rho^0}{\sqrt{2}} + \frac{\omega}{\sqrt{2}} & K^{*0} & D^{*-} \\
 & & & \\
  K^{*-} & \bar K^{*0} & \phi & D_s^{*-} \\
  & & & \\
D^{*0} & D^{*+} & D_s^{*+} & J/\psi \\ 
\end{array}
\right)_\mu \ .
\end{equation}

The interaction of ${\cal L}_{III}$ provides a contact term which comes from 
$[V_\mu,V_\nu][V_\mu,V_\nu]$
\begin{equation}
{\cal L}^{(c)}_{III}=\frac{g^2}{2}\langle V_\mu V_\nu V^\mu V^\nu-V_\nu V_\mu
V^\mu V^\nu\rangle\ ,
\label{lcont}
\end{equation}
as well as to a three 
vector vertex from 
\begin{equation}
{\cal L}^{(3V)}_{III}=ig\langle (\partial_\mu V_\nu -\partial_\nu V_\mu) V^\mu V^\nu\rangle
=ig\langle (V^\mu\partial_\nu V_\mu -\partial_\nu V_\mu
V^\mu) V^\nu\rangle \ .
\label{l3Vsimp}
\end{equation}

It is worth recalling the analogy with the coupling of vectors to
 pseudoscalars given in the same formalism by  
\begin{equation}
{\cal L}_{VPP}= -ig ~\langle [P,\partial_{\mu}P]V^{\mu}\rangle,
\label{lagrVpp}
\end{equation}
where $P$ is the SU(4) matrix of the pseudoscalar fields,
\begin{equation}
P = \left(
\begin{array}{cccc}
\frac{\pi^0}{\sqrt{2}}+\frac{\eta_8}{\sqrt{6}} +\frac{\tilde{\eta}_c}{\sqrt{12}}+\frac{\tilde{\eta}_c'}{\sqrt{4}}  &\pi^+     & K^{+}    &\bar{D}^{0}\\
\pi^-      & -\frac{\pi^0}{\sqrt{2}}+\frac{\eta_8}{\sqrt{6}}+\frac{\tilde{\eta}_c}{\sqrt{12}}+\frac{\tilde{\eta}_c'}{\sqrt{4}}& K^{0}    &D^{-}\\
K^{-}      & \bar{K}^{0}       &\frac{-2\eta_8}{\sqrt{6}}+\frac{\tilde{\eta}_c}{\sqrt{12}}+\frac{\tilde{\eta}_c'}{\sqrt{4}}                            &D^{-}_s\\
D^{0}&D^{+}&D^{+}_s&-\frac{3\tilde{\eta}_c}{\sqrt{12}}+\frac{\tilde{\eta}_c'}{\sqrt{4}}\\
\end{array}
\right) \ .
\end{equation}
where $\tilde{\eta}_c$ stands for the SU(3)
singlet of the 15th SU(4) representation and we denote
$\tilde{\eta}'_c$ for the singlet of SU(4) (see quark content in \cite{wuprc}). The physical $\eta_c$ can be written as \cite{wuprc}
\begin{equation}
\eta_c = \frac{1}{2}(-\sqrt{3}\tilde{\eta}_c+\tilde{\eta}'_c)\ .
\end{equation}

\begin{figure}[tb]
\epsfig{file=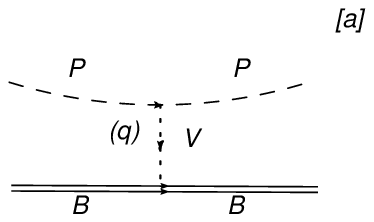, width=7cm} \epsfig{file=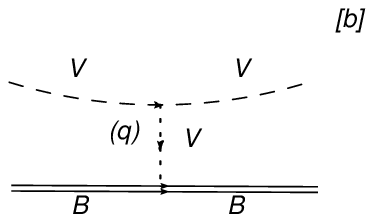, width=7cm}
\caption{Diagrams obtained in the effective chiral Lagrangians for interaction
of pseudoscalar [a] or vector [b] mesons with the octet or decuplet of baryons.}%
\label{f1}%
\end{figure}

The philosophy of the local hidden gauge in the meson-baryon sector is that the interaction is driven by the exchange of vector mesons, as depicted in Fig. \ref{f1}.
Eqs. \eqref{l3Vsimp} and \eqref{lagrVpp} provide the upper vertex of these Feyman diagrams. It was shown in \cite{angelsvec} that the vertices of Eq. (\ref{l3Vsimp}) and Eq. (\ref{lagrVpp}) give rise to the same expression in the limit of small three momenta of the vector mesons compared to their mass, a limit which is also taken in our calculations.  This makes the work technically easy and it allows the use of many previous results. 

The lower vertex when the baryons belong to the octet of SU(3) is given in terms of the Lagrangian \cite{Klingl:1997kf,Palomar:2002hk} 
\begin{equation}
{\cal L}_{BBV} =
\frac{g}{2}\left(\langle\bar{B}\gamma_{\mu}[V^{\mu},B]\rangle+\langle\bar{B}\gamma_{\mu}B\rangle \langle V^{\mu}\rangle \right),
\label{lagr82}
\end{equation}
where $B$ is now the SU(3) matrix of the baryon octet \cite{Eck95,Be95}. Similarly,
one has also a Lagrangian for the coupling of the vector mesons to the baryons
of the decuplet, which can be found in \cite{manohar}.

In the charm sector the lower vertex $VBB$ does not have such a simple representation as in SU(3) and in practice one evaluates the matrix elements using SU(4) symmetry by means of Clebsch-Gordan coefficients and reduced matrix elements. This is done in \cite{wuprl,wuprc} (a discussion on the accuracy of the SU(4) symmetry is done there). Since the 20 representation for baryon states of $3/2^+$ is not considered there, we must consider these matrix elements here too. Once again one uses SU(4) symmetry for this vertex to evaluate the matrix elements, as done in \cite{wuprl,wuprc}. Alternatively, one can use results of SU(3) symmetry substituting a $s$ quark by a $c$ quark, or make evaluations using wave functions of the quark model \cite{close}, substituting the $s$ quark by the $c$ quark.

The $\gamma^\mu$ matrix of the $VBB$ vertex (see Eq. \eqref{lagr82}) gets simplified in the approach, where we neglect the three momenta versus the mass of the particles (in this case the baryon). Thus, only the $\gamma^0$ becomes relevant, which can be taken as unity within the baryon states of positive energy that we consider. Then the transition potential corresponding to the diagram of Fig. \ref{f1}(b) is given by
\begin{equation}
V_{i j}= - C_{i j} \, \frac{1}{4 f^2} \, (k^0 + k'^0)~ \vec{\epsilon} \ \vec{\epsilon
}\ ',
\label{eq:kernel}
\end{equation}
 where $k^0, k'^0$ are the energies of the incoming and outgoing vector mesons, and $C_{ij}$ numerical coefficients evaluated as described above. The expression is the same for the pseudoscalar baryon matrix elements for the same quark content of pseudoscalar and vector mesons, omitting the $\vec{\epsilon}~\vec{\epsilon }~'$ factor. 

The scattering matrix is evaluated by solving the coupled channels Bethe-Salpeter equation in the on shell factorization approach of \cite{angels,ollerulf,Nieves:1999bx}
\begin{equation}
T = [1 - V \, G]^{-1}\, V,
\label{eq:Bethe}
\end{equation} 
with $G$ being the loop function of a meson and a baryon, which we calculate in dimensional regularization using the formula of \cite{ollerulf} and similar values for the subtraction constants.

The iteration of diagrams produced by the Bethe Salpeter equation in the case of the vector mesons keeps the $\vec{\epsilon}~\vec{\epsilon }~'$ factor in each of the terms. Hence, the factor $\vec{\epsilon}~\vec{\epsilon }~'$ appearing in the potential $V$ factorizes also in the $T$ matrix for the external vector mesons. A consequence of this is that the interaction is spin independent and one finds degenerate states having $J^P=1/2^-$ and $J^P=3/2^-$.

In the present work, in the spirit of the heavy quark symmetry, we shall include in the coupled channels dynamics, the pseudoscalars, vectors, baryons of spin $J=1/2$ and baryons of $J=3/2$ using the matrices of Eqs. (\ref{eq:ji11}$-$\ref{eq:ji53}).

\section{Evaluation of the HQSS LEC's in the local hidden gauge approach}

Let us examine first the $I=1/2$ sector. As an example let us take $\bar{D} \Lambda_c \to \bar{D} \Lambda_c$ and $\bar{D}^* \Lambda_c \to \bar{D}^* \Lambda_c$. These two interactions are equal as we discussed. This is in agreement with the general HQSS constraints explicited in Eq. \eqref{eq:ji11} for $J=1/2$ and $I=1/2$, where both matrix elements are equal to the LEC's $\mu_2$, and it is also consistent with the diagonal $\bar{D}^* \Lambda_c$ entry in Eq. \eqref{eq:ji31} ($J=3/2,~I=1/2$). So we see that the HQSS is respected there by the local hidden gauge results. In addition the interactions of $\bar{D} \Sigma_c \to \bar{D} \Sigma_c$ and $\bar{D}^* \Sigma_c \to \bar{D}^* \Sigma_c$ are also equal. This does not contradict Eqs. \eqref{eq:ji11} and \eqref{eq:ji31}, it simply forces
\begin{equation}
\frac{1}{3} (2 \lambda_2 + \mu_3) = \frac{1}{9} (2 \lambda_2 + 7 \mu_3),
\end{equation}
which has as a solution,
\begin{equation}
\lambda_2 = \mu_3.
\end{equation}
This has as a consequence that the matrix element of $\bar{D}^* \Sigma_c^*  \to \bar{D}^* \Sigma_c^* $ is also equal to $\lambda_2$. The evaluation of this later matrix element using SU(4) Clebsch-Gordan coefficients also tells us that this matrix element is the same as the one of $\bar{D}^* \Sigma_c \to \bar{D}^* \Sigma_c$. Once again we can see that the constraints of HQSS are fulfilled by the hidden gauge formalism, only that it gives us $\lambda_2 = \mu_3$, which is a result different to the one obtained in the approach of \cite{Garcia-Recio:2013gaa}.

\begin{figure}[tb]
\epsfig{file=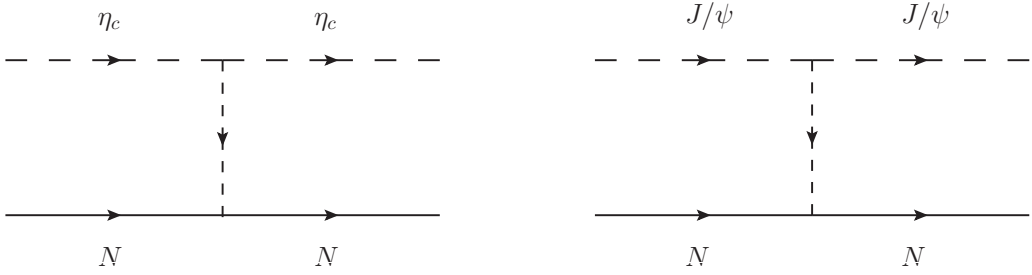, width=14cm}
\caption{Diagrams for the $\eta_c N \to \eta_c N$ and $J/\psi N \to J/\psi N$ interactions. No vector meson exchanged is allowed.}%
\label{fig:mu1}%
\end{figure}

Let us look at the coefficients $\mu_1$. It is related to the $\eta_c N \to \eta_c N$ or $J/\psi N \to J/\psi N$ matrix elements. In this case with the diagram of Fig. \ref{fig:mu1},
 since $\eta_c$ or $J/\psi$ have $c \bar{c}$, there is no vector that can be exchanged in Fig. \ref{fig:mu1} and hence this leads to
\begin{equation}
\mu_1 = 0.
\end{equation}

\begin{figure}[tb]
\epsfig{file=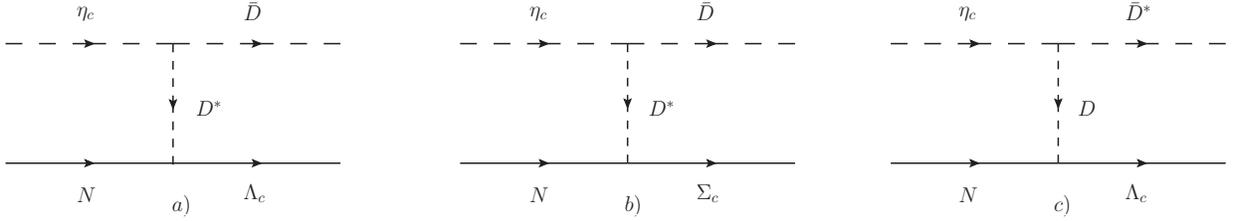, width=16.5cm}
\caption{Diagrams for the $\eta_c N \to \bar{D} \Lambda_c, \bar{D} \Sigma_c, \bar{D}^* \Lambda_c$ interaction.}%
\label{fig:mu123}%
\end{figure}

This also occurs in the approach of \cite{Garcia-Recio:2013gaa} and it is a consequence of the OZI rule, that is implemented in both schemes. Let us now look at the $\mu_{12}$ parameter. This enters in $\eta_c N \to \bar{D} \Lambda_c$ transition which is depicted the diagram of Fig. \ref{fig:mu123} a).
Within the hidden gauge model, the diagram forces the exchange of a $D^*$ and is subleading in the $m_Q$ counting (${\cal O}(m_Q ^{-2})$). In the limit of $m_Q \to \infty$ this term would vanish. We, however, keep it and take it from \cite{wuprl, wuprc}. Yet, because it is subleading we shall not expect the LO HQSS restrictions to hold. We also evaluate the diagram of Fig. \ref{fig:mu123} b),
and using again SU(4) symmetry for the $D^* N \Sigma_c$ vertex (see \cite{wuprl, wuprc}) we find that
\begin{equation}
\frac{\mu_{13}}{2} = -\frac{\mu_{12}}{2} \ \Rightarrow \ \mu_{13} = -\mu_{12},
\end{equation}
which also occurs in \cite{Garcia-Recio:2013gaa}.

As to the transition from $\eta_c N \to \bar{D}^* \Lambda_c$, they are mediated by the exchange of a $D$ meson, see Fig. \ref{fig:mu123} c).
This term is doubly suppressed because of the $D$ propagator and because of the Yukawa coupling, $\vec{\sigma} \cdot \vec{q}$, in $D N \Lambda_c$ vertex, where the three momentum is small compared with $m_D$. In Eq. \eqref{eq:ji11} we see that this term is proportional to $\mu_{12}$, showing again that the LO HQSS constraints does not hold for these subleading terms in the $m_Q$ counting. In practice keeping this term, and those for $\eta_c N \to \bar{D}^* \Sigma_c, ~\bar{D}^* \Sigma_c^*$ or ignoring them has no practical repercussion on the final results.

\subsection{Transition from $D,\ D^*$}

\begin{figure}[tb]
\epsfig{file=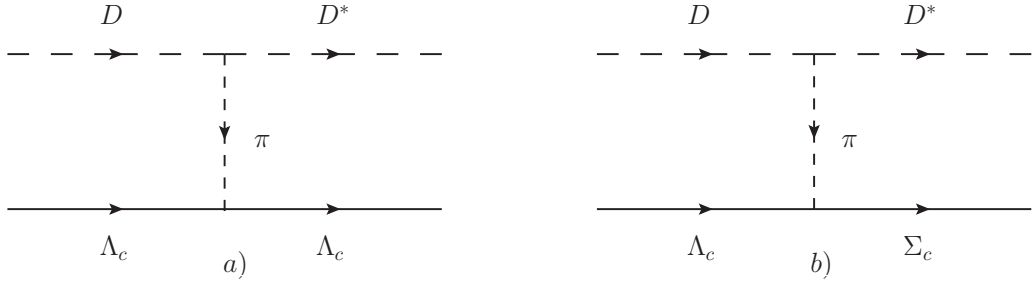, width=14cm}
\caption{Diagrams for the pion exchange in the transition of $D,\ D^*$.}%
\label{fig:pionex}%
\end{figure}

With the dynamics of the local hidden gauge approach only the pion exchange in the t-channel is allowed in this case, see Fig. \ref{fig:pionex}.
The $D \Lambda_c \to D^* \Lambda_c$ transition is zero because the $\pi$ exchange is zero in $\pi \Lambda_c \Lambda_c$ vertex. This agrees with the result of the matrix of Eqs. \eqref{eq:ji11}, \eqref{eq:ji31} and \eqref{eq:ji51}. However the transition $D \Lambda_c \to D^* \Sigma_c$ is not null and we evaluate it here.

The $\pi \Lambda_c \Sigma_c$ vertex can be obtained by analogy to the $\pi \Lambda \Sigma$ vertex in SU(3) (exchanging $c$ and $s$ quark) and using the Lagrangian,
\begin{equation}
{\cal L} = \frac{1}{2} D \langle \bar{B} \gamma^\mu \gamma_5 \{u_\mu,B\} \rangle + \frac{1}{2} F \langle \bar{B} \gamma^\mu \gamma_5 [u_\mu,B] \rangle,
\end{equation}
where $u_\mu = i u^\dagger \partial_\mu U u^\dagger$, $u^2 = U = e^{i \sqrt{2}\phi /f}$ with $D = 0.80,\ F = 0.46$ from \cite{Borasoy:1998pe}. The $D D^* \pi$ vertex is evaluated from Eq. \eqref{lagrVpp}. We find at the end projecting over s-wave,
\begin{equation}
-i t = \frac{1}{\sqrt{6}}\ \frac{M_V}{2f}\ \frac{2}{5}\ \frac{D+F}{2f}\ \vec{q}\,^2\ \vec{\sigma} \cdot \vec{\epsilon}\ \frac{i}{q^{0\ 2} - \vec{q}\,^2 - m_\pi^2},\label{eq:tpiexcha}
\end{equation}
with $\vec{q}$ the momentum transfer.

One can also prove that the matrix element of $\vec{\sigma} \cdot \vec{\epsilon}$ is $\sqrt{3}$ \cite{javier}. If we compare this contribution of this diagram with that of the $D \Lambda_c \to D \Lambda_c$ transition from \cite{wuprl, wuprc}, we find a contribution of the order of 7\%. If one looks at diagonal matrix elements in the final scattering T-matrix, the non diagonal terms of the transition potentials come squared and then we can safely neglect this contribution. Thus we take
\begin{equation}
\mu_{23} = 0.
\end{equation}
Note that the transitions $\bar{D} \Sigma_c \to \bar{D}^* \Sigma_c,\ \bar{D}^* \Sigma_c^*$ also require the pion exchange and should be taken zero. This is consistent with the matrix of Eq. \eqref{eq:ji11} since these matrix elements are proportional to $\lambda_2 - \mu_3$ but we saw before that $\lambda_2 = \mu_3$.

When evaluating the pion exchange mechanism in the $VB \to VB$ transition one has to consider the equivalent contact term that in the case of $\gamma N \to \pi N$ scattering is known as the Kroll Ruderman term. Explicit expressions to obtain it can be found from \cite{javier,kanchan1,kanchan2,kanchan3} and is of the same order of magnitude as the pion exchange term, with usually destructive interference. We do not need to evaluate it explicitly here because the important point is that, as Eq. \eqref{eq:tpiexcha}, it is of order ${\cal O}(1)$ in the $m_Q$ counting for the field theoretical potential, which implies ${\cal O}(m_Q^{-1})$ for the ordinary potential of Quantum Mechanics, as we shall see in the next section.

With this exercise we have proved that the dynamics of the local hidden gauge approach is fully consistent with the HQSS requirements for the matrix of Eq. \eqref{eq:ji11}. The values for the parameters that we obtain from \cite{wuprl, wuprc}, together with those determined here, are
\begin{equation}
\begin{split}
\mu_2 &= \frac{1}{4f^2} (k^0 + k'^0),\\
\mu_3 &= -\frac{1}{4f^2} (k^0 + k'^0),\\
\mu_{12} &= -\sqrt{6}\ \frac{m_\rho^2}{p^2_{D^*} - m^2_{D^*}}\ \frac{1}{4f^2}\ (k^0 + k'^0),\\
\mu_1 &= 0,\\
\mu_{23} &= 0,\\
\lambda_2 &= \mu_3,\\
\mu_{13} &= -\mu_{12}.\label{eq:ji11fi}
\end{split}
\end{equation}
$\mu_{12}$ is small, of the order of 15\%. But we keep it since this term is the only one that allows the scattering $\eta_c N \to \eta_c N$ ($J/\psi N \to J/\psi N$) through intermediate inelastic states.

\begin{figure}[tb]
\epsfig{file=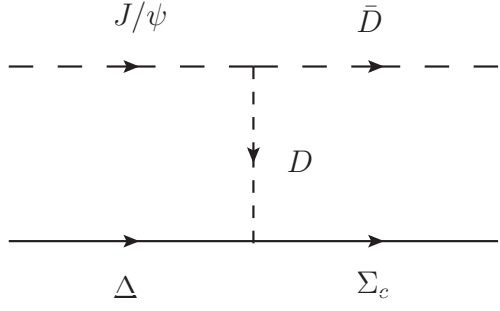, width=7cm}
\caption{Diagrams for the $J/\psi \Delta \to \bar{D} \Sigma_c$ interaction.}%
\label{fig:lam12}%
\end{figure}

The matrix of Eq. \eqref{eq:ji13} for $J=1/2$ and $I=3/2$ is equally analyzed. We find
\begin{equation}
\lambda_1 = 0.
\end{equation}
Then $\lambda_{12}$ is also suppressed since it requires again the exchange of a $D$ meson, see Fig. \ref{fig:lam12}.
Once again, since the $\bar{D} \Sigma_c \to \bar{D} \Sigma_c$ transition is equivalent to $\bar{D}^* \Sigma_c \to \bar{D}^* \Sigma_c$. This implies that
\begin{equation}
\frac{1}{3} (2 \lambda_2 + \mu_3) = \frac{1}{9} (2 \lambda_2 + 7 \mu_3),
\end{equation}
from where we conclude again that
\begin{equation}
\lambda_2 = \mu_3.
\end{equation}
Once again the $\bar{D} \Sigma_c \to \bar{D}^* \Sigma_c,\ \bar{D} \Sigma_c^*$ transitions involve pion exchange and we find them negligible, which is compatible with the HQSS requirement since $\mu_3 - \lambda_2 = 0$. The values that we obtain with this isospin combination are
\begin{equation}
\begin{split}
\lambda_{12} &= 3\sqrt{3}\ \frac{m_\rho^2}{p^2_{D^*} - m^2_{D^*}}\ \frac{1}{4f^2}\ (k^0 + k'^0),\\
\mu_3 &= 2 \frac{1}{4f^2} (k^0 + k'^0),\\
\lambda_2 &= \mu_3,\\
\lambda_1 &= 0. \label{eq:ji13fi}
\end{split}
\end{equation}

For Eq. 	\eqref{eq:ji31} ($J=3/2, I=1/2$) since our interaction is spin independent for $P B \to P B$ and of the type $\vec{\epsilon}~ \vec{\epsilon}~'$ for $VB \to VB$, then the coefficients are the same as those for Eq. \eqref{eq:ji11} ($J=1/2,~ I=1/2$), given in Eq. \eqref{eq:ji11fi}.

The same can be said for the matrix of Eq. \eqref{eq:ji33} with respect to the one of Eq. \eqref{eq:ji13}, which are given in Eq. \eqref{eq:ji13fi}.

As to Eq. \eqref{eq:ji51}, once again $\bar{D}^* \Sigma_c^* \to \bar{D}^* \Sigma_c^*$ has the same matrix element as $\bar{D}^* \Sigma_c^* \to \bar{D}^* \Sigma_c^*$ of Eq. \eqref{eq:ji11} and, indeed, since $\lambda_2 = \mu_3,~ \frac{1}{9} (\lambda_2 + 8 \mu_3) = \lambda_2$, which is given in Eq. \eqref{eq:ji11fi}.

Finally in Eq. \eqref{eq:ji53} $\lambda_1 = 0$ for us and $\lambda_2,~ \lambda_{12}$ the same as those given in Eq. \eqref{eq:ji13fi}($I=3/2$).

\section{Heavy quark spin symmetry in the SU(4) extended hidden gauge approach}

The origin of the heavy quark spin symmetry in this case is easy to trace. The $PB \to PB$ transitions have no spin dependence. Also, under the approximation that $\vec{q}/M_D,\ \vec{q}/M_{D^*}$ are negligible (consistent with the heavy quark symmetry), the $VB \to VB$ interaction has the trivial $\vec{\epsilon}~ \vec{\epsilon}~'$ dependence and no spin dependence of the baryons which also leads to spin independence.
Also up to the trivial $\vec{\epsilon}~ \vec{\epsilon}~'$ factor the $\bar{D}^* B \to \bar{D}^* B$ interaction is the same as the one for $\bar{D} B \to \bar{D} B$.

Heavy quark symmetry also implies that in leading order the potential is independent of the flavour of the heavy quarks in the limit of $m_Q \to \infty$. This is also accomplished by the dynamics of the local hidden gauge approach. This might be surprising since, as seen in Eq. \eqref{eq:kernel}, the potential goes like the sum of the energies of the mesons. This obviously grows from the strange to the charm and then to the bottom sector. However, this is the potential in the field theoretical approach. To have an idea of the strength of the interaction one has to converts this into the ordinary potential that appears in the Schr\"{o}dinger equation of Quantum Mechanics. This is done in \cite{gamer} (Eq. (68) of that reference).

Because of the normalization of the field ($\sqrt{\frac{2M_B}{2E_B}}$ for baryons and $\frac{1}{\sqrt{2 \omega_M}}$ for mesons) we have
\begin{equation}
V^{FT} = \frac{32 \pi^3}{2 M_B} \sqrt{s}\ \mu\ v^{QM}, \text{   (meson-baryon)}
\end{equation}
where $\mu$ is the reduced mass of the system, $M_B m_D / (M_B + m_D)$. Considering that the leading potentials go as $k^0 + k'^0$, we find that
\begin{equation}
v^{QM} \sim \frac{(1 + \frac{M_B}{m_D})^2}{\frac{1}{2} + \frac{M_B}{m_D} + \frac{M_B^2}{2m_D^2}} \sim {\cal O}(m_Q^0),
\end{equation}
which goes as ${\cal O} (1)$ in powers of $m_D$, both if $M_B$ is a nucleon or if $M_B$ is one charmed baryon.

Incidentally, in the case of meson-meson interaction the formula is
\begin{equation}
V^{FT} = 32 \pi^3 \sqrt{s}\ \mu\ v^{QM}, \text{   (meson-meson)}
\end{equation}
Here the interaction in field theory goes as $V^{FT} \sim (k^0 + k'^0)(p^0 + p'^0)$ since both meson lines are linked by a vector meson and we have the $(k^0 + k'^0)$ in each vertex. In this case we also find immediately that $V^{QM} \sim {\cal O} (1)$. This means that the extrapolation of the rules of the hidden gauge to SU(4) or even to the bottom sector are strictly fulfilling the rules of LO HQSS. This would justify the work of \cite{wuzou}, where a direct extrapolation of the work \cite{wuprl, wuprc} to the beauty sector is done, or of \cite{GarciaRecio:2012db}, where a direct extrapolation of the Weinberg-Tomozawa term is taken also in the beauty sector.

\section{Results and discussion}

We use the Bethe-Salpeter equation of Eq. \eqref{eq:Bethe} in coupled channels to evaluate the scattering amplitudes. We need the $G$ function, loop function of meson-baryon interaction, for which we take the usual dimensional regularization formula \cite{ollerulf} 
\begin{eqnarray}
G(s) &=&i \int\frac{d^{4}q}{(2\pi)^{4}}\frac{2M_{B}}{(P-q)^{2}-M^{2}_{B}+i\varepsilon}\,\frac{1}{q^{2}-M^{2}_{P}+i\varepsilon},\\
&=&\frac{2M_{B}}{16\pi^2}\big\{a_{\mu}+\textmd{ln}\frac{M^{2}_{B}}{\mu^{2}}+\frac{M^{2}_{P}-M^{2}_{B}+s}{2s}\textmd{ln}\frac{M^{2}_{P}}{M^{2}_{B}}\nonumber\\
&&+\frac{q_{cm}}{\sqrt{s}}\big[\textmd{ln}(s-(M^{2}_{B}-M^{2}_{P})+2q_{cm}\sqrt{s})+\textmd{ln}(s+(M^{2}_{B}-M^{2}_{P})+2q_{cm}\sqrt{s})\nonumber\\
&&-\textmd{ln}(-s-(M^{2}_{B}-M^{2}_{P})+2q_{cm}\sqrt{s})-\textmd{ln}(-s+(M^{2}_{B}-M^{2}_{P})+2q_{cm}\sqrt{s})\big]\big\}\ ,\label{eq:G}
\end{eqnarray}
where $q$ is the four-momentum of the meson, $q_{cm}$ the three-momentum of the particle in the center mass frame,  and $P$ is the total four-momentum of the meson and the baryon, thus, $s=P^2$. This formula avoids an undesired behaviour at large energies when one uses a cut off method with a small cut off \cite{Guo:2005wp}. As done in \cite{wuprl,wuprc}, we take $\mu=1000 \mev,~a(\mu)=-2.3$ for the parameters in Eq. \eqref{eq:G}, which are the only free parameters in our present study. We solve the Bethe-Salpeter equation of Eq. \eqref{eq:Bethe} in coupled channels and look for poles in the second Riemann sheet when there are open channels, or in the first Riemann sheet when one has stable bound states (see \cite{wuprc,luis} for details).

Let $\sqrt{s_p}$ be the complex energy where a pole appears. Close to a pole the amplitude behaves as
\begin{equation}
T_{ij}=\frac{g_{i}g_{j}}{\sqrt{s}-\sqrt{s_p}}\ .\label{eq:tgigj}
\end{equation}
where $g_i$ is the coupling of the resonance to the $i$ channel. As one can see in Eq. \eqref{eq:tgigj}, $g_{i}g_{j}$ is the residue of $T_{ij}$ at the pole. For a diagonal transitions we have
\begin{equation}
g_{i}^{2}=\lim_{\sqrt{s}\rightarrow
\sqrt{s_p}}~T_{ii}\,(\sqrt{s}-\sqrt{s_p}).\label{eq:coup}
\end{equation}
The determination of the couplings gives as an idea of the structure of the states found, since according to \cite{gamer,junko2}, the couplings are related to the wave function at the origin for each channel.

\begin{figure}
\centering
%\subfigure[\ The squared amplitudes of three channels except for $J/\psi \Delta$ channel.]{\label{fig:i32a}\includegraphics[scale=0.6]{tsqrt1a_1234.eps}}
\subfigure[]{\label{fig:i32a}\includegraphics[scale=0.6]{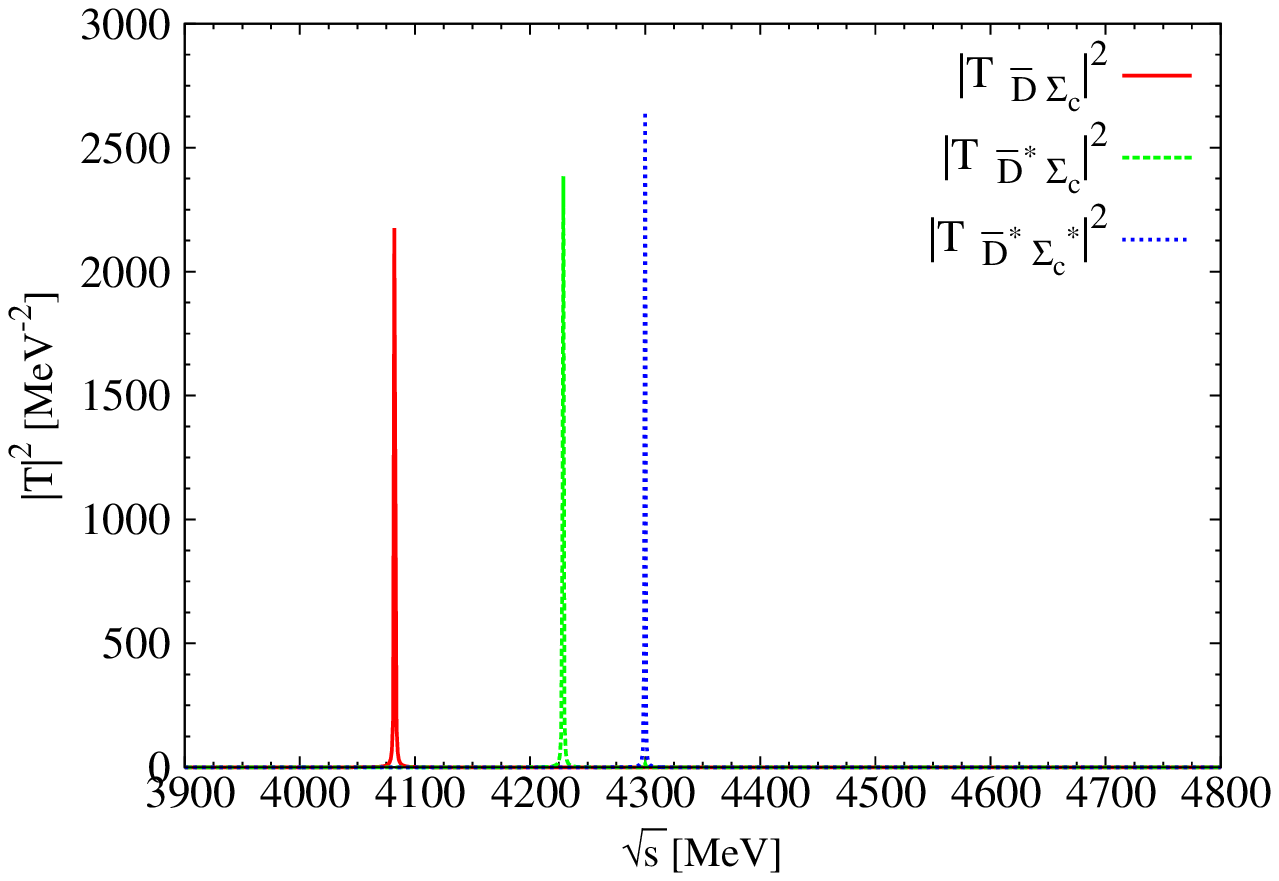}}
\subfigure[]{\label{fig:i32b}\includegraphics[scale=0.8]{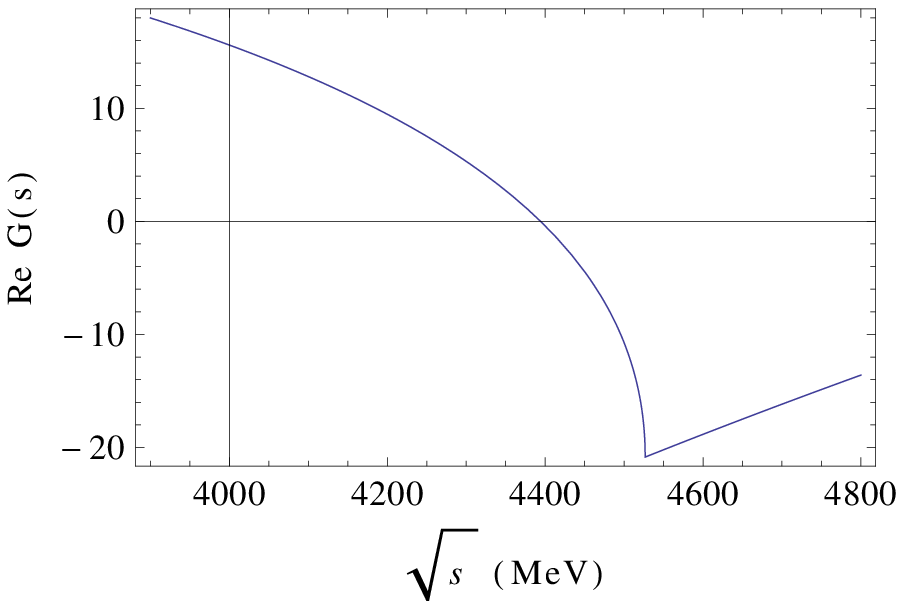}}
\caption{The results of the $J=1/2,~I=3/2$ sector. (a): The squared amplitudes of three channels except for $J/\psi \Delta$ channel. (b): The real parts of $G$ function in $\bar{D}^* \Sigma_c^*$ channel.}
\end{figure}

Let us begin with the $J=1/2,~I=3/2$ sector. We can see in the Eq. \eqref{eq:ji13fi} that the large potentials are repulsive. So, we should not expect any bound states or resonances. Yet, technically we find bound states in the first Riemann sheet, as one can see in Fig. \ref{fig:i32a} for different channels. 
However, inspection of the energies tell us that these are states bound by about 250 MeV, a large number for our intuition, even more when we started from a repulsive potential. The reason for this, which forces us to reject these poles on physical grounds, is that the $G$ function below threshold turns out to be positive for large binding energies (see Fig. \ref{fig:i32b} and discussions in \cite{wuzou}), contradicting what we would have for the $G$ function evaluated with any cut off, or in Quantum Mechanics with a given range. These poles are then discarded and, thus, we do not find bound states or resonances in $I=3/2$ in our approach. 

The WT extended model of Ref.\cite{Garcia-Recio:2013gaa} predicts $\mu_3=-2$, which leads to some attractive interactions in the space generated by ${\bar D}^* \Sigma_c$ , ${\bar D} \Sigma^*_c$ and ${\bar D}^* \Sigma^*_c$. These give rise to three odd parity $\Delta-$like resonances (two with spin 1/2 and one with spin 3/2) with masses around 4 GeV. In addition, two other states show up as cusps very close to the $\Delta J/\psi$ threshold, and their real existence would be unclear.

\begin{figure}[tb]
\epsfig{file=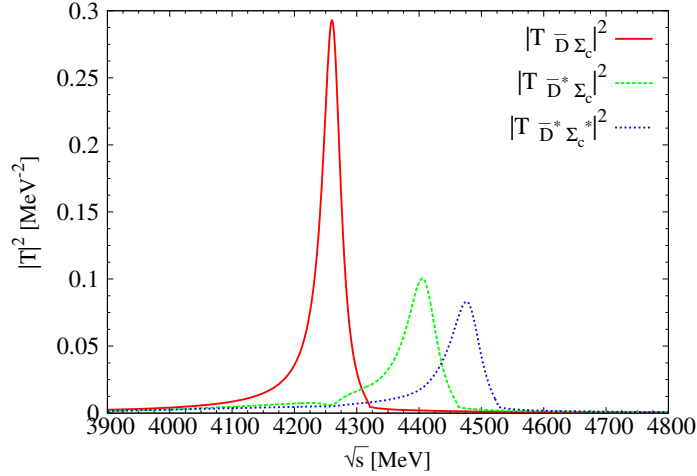, width=9cm}
\caption{The squared amplitudes of the $J=1/2,~I=1/2$ sector.}\label{fig:res11}%
\end{figure}

Our results for the $J=1/2,~I=1/2$ sector are shown in Fig. \ref{fig:res11}.
From the squared amplitudes of $|T|^2$, we can find three clear peaks with non zero width around the energy range $4200 \sim 4500 \mev$, which are not far away below the thresholds of $\bar{D} \Sigma_c, ~\bar{D}^* \Sigma_c, ~\bar{D}^* \Sigma_c^*$ respectively. The relatively small width of about $40\mev$ of these states allows to distinguish them clearly. We have checked that in the energy ranges where these peaks appear, the real parts of the loop function $G$, Eq. \eqref{eq:G}, are negative in these channels. Thus these peaks are acceptable as physical ones. Then, because of the non zero width, we look for the poles corresponding to these peaks in the second Riemann sheet, and find the poles at $(4261.87+i17.84)\mev, ~(4410.13+i29.44)\mev, ~(4481.35+i28.91)\mev$. The couplings to the various coupled channels for these poles are given in Table \ref{tab:cou11}.
\begin{table}[ht]
     \renewcommand{\arraystretch}{1.2}
\centering
\caption{The coupling constants of all channels corresponded certain poles in the $J=1/2,~I=1/2$ sector.} \label{tab:cou11}
\begin{tabular}{cccc cccc}
\hline\hline
\multicolumn{2}{c}{$4261.87+i17.84$}  \\
\hline
   & $\eta_c N$ & $J/\psi N$ & $\bar{D} \Lambda_c$ & $\bar{D} \Sigma_c$ & $\bar{D}^* \Lambda_c$ & $\bar{D}^* \Sigma_c$ & $\bar{D}^* \Sigma_c^*$  \\
\hline
$g_i$ & $1.04+i0.05$ & $0.76-i0.08$ & $0.02-i0.02$ & $3.12-i0.25$ & $0.14-i0.48$ & $0.33-i0.68$ & $0.16-i0.28$  \\
$|g_i|$ & $1.05$ & $0.76$ & $0.02$ & $3.13$ & $0.50$ & $0.75$ & $0.32$  \\
\hline
\multicolumn{2}{c}{$4410.13+i29.44$}  \\
\hline
    & $\eta_c N$ & $J/\psi N$ & $\bar{D} \Lambda_c$ & $\bar{D} \Sigma_c$ & $\bar{D}^* \Lambda_c$ & $\bar{D}^* \Sigma_c$ & $\bar{D}^* \Sigma_c^*$  \\
\hline
$g_i$ & $0.34+i0.16$ & $1.43-0.12$ & $0.15-i0.10$ & $0.20-i0.05$ & $0.17-i0.11$ & $3.05-i0.54$ & $0.07-i0.51$  \\
$|g_i|$ & $0.38$ & $1.44$ & $0.18$ & $0.20$ & $0.20$ & $3.10$ & $0.51$  \\
\hline
\multicolumn{2}{c}{$4481.35+i28.91$}  \\
\hline
   & $\eta_c N$ & $J/\psi N$ & $\bar{D} \Lambda_c$ & $\bar{D} \Sigma_c$ & $\bar{D}^* \Lambda_c$ & $\bar{D}^* \Sigma_c$ & $\bar{D}^* \Sigma_c^*$  \\
\hline
$g_i$ & $1.15-i0.04$ & $0.72+i0.03$ & $0.18-i0.08$ & $0.10-i0.03$ & $0.09-i0.08$ & $0.09-i0.06$ & $2.88-i0.57$  \\
$|g_i|$ & $1.15$ & $0.72$ & $0.19$ & $0.10$ & $0.12$ & $0.11$ & $2.93$  \\
\hline
\end{tabular}
\end{table}
From Table \ref{tab:cou11} we can see that the first pole, $(4261.87+i17.84)\mev$, couples mostly to $\bar{D} \Sigma_c$. It could be considered like a $\bar{D} \Sigma_c$ bound state which, however, decays into  the open channels $\eta_c N$ and $J/\psi N$. The $\bar{D} \Sigma_c$ threshold is at $4320.8\mev$ and, thus, the $\bar{D} \Sigma_c$ state is bound  by about $58\mev$. The second pole couples most strongly to $\bar{D}^* \Sigma_c$. In this channel the threshold is $4462.2\mev$ and thus we have a state bound by about $52\mev$, much in line with what one expects from heavy quark symmetry comparing this with the former state. This state decays mostly into the $\eta_c N$ and $J/\psi N$ channels again. These two states correspond to those reported in \cite{wuprl,wuprc}. In our work, we get one more new baryon state, $(4481.35+i28.91)\mev$, with total momentum $J=1/2$, which couples mostly to $\bar{D}^* \Sigma_c^*$. Since in \cite{wuprl,wuprc} one did not include the baryons of $J^P=3/2^+$, their consideration here leads to a new resonance. The threshold for the $\bar{D}^* \Sigma_c^*$ channel is $4526.7\mev$ and, hence, the state can be considered as a $\bar{D}^* \Sigma_c^*$ bound state by about $46\mev$, which decays mostly in $\eta_c N$ and $J/\psi N$.

\begin{figure}[tb]
\epsfig{file=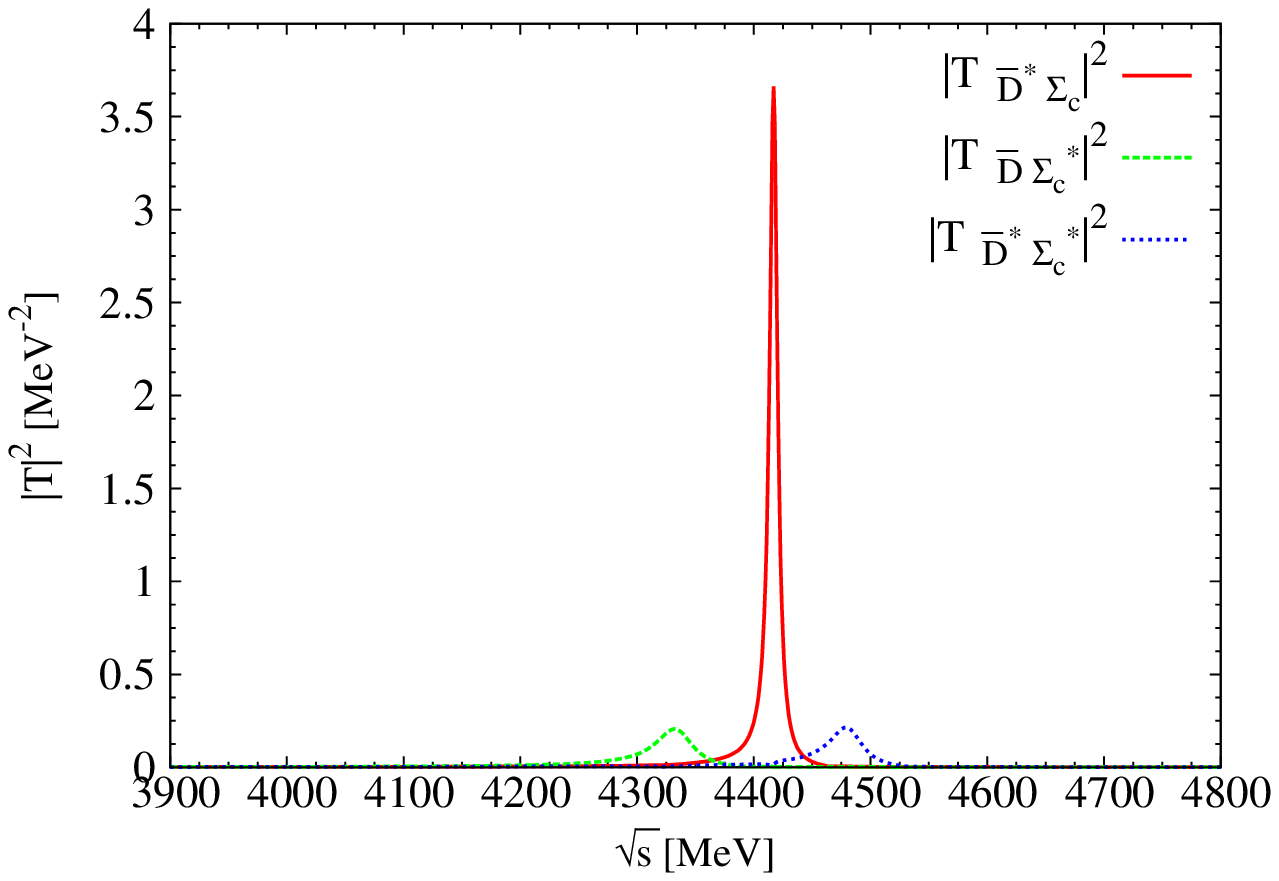, width=8cm}
\epsfig{file=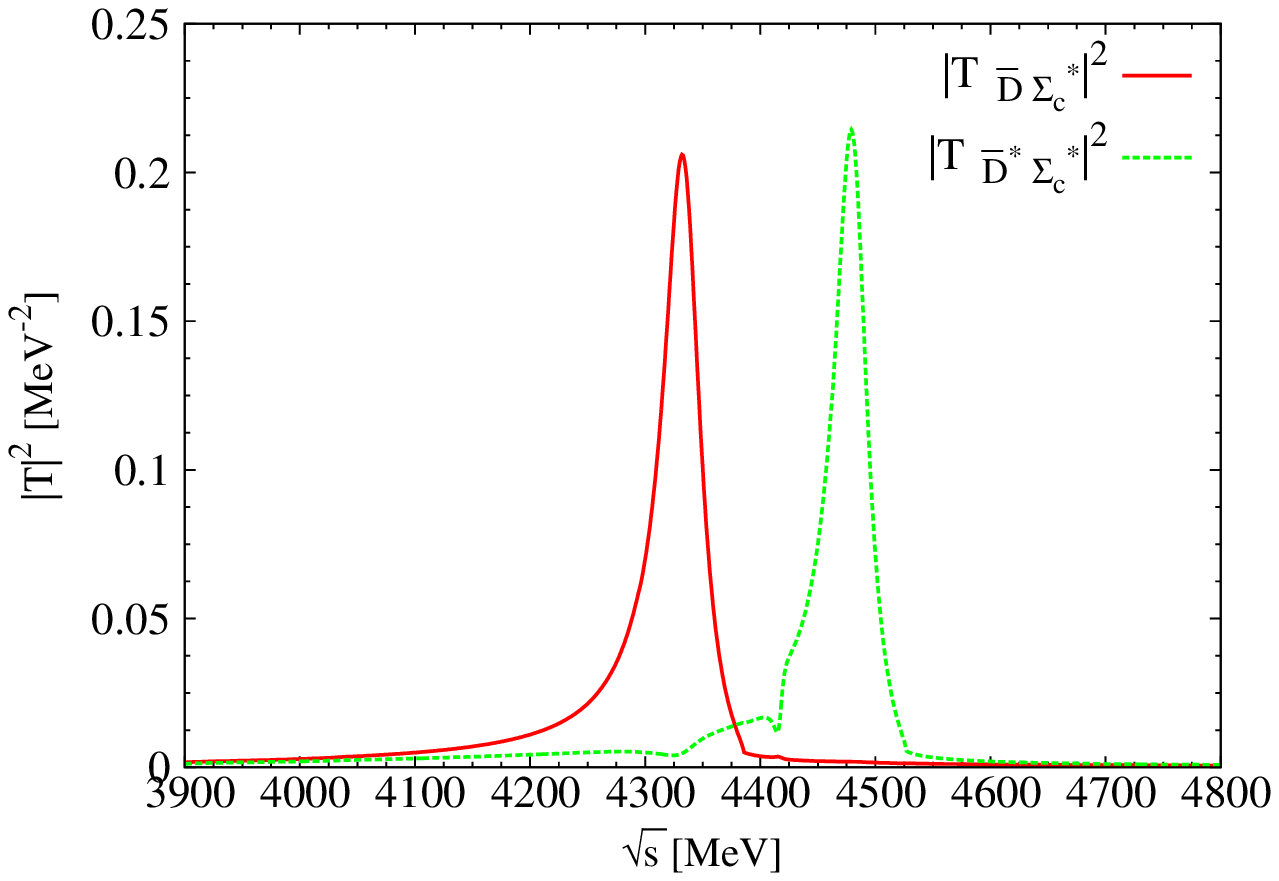, width=8cm}
\caption{The results of $|T|^2$ for the $J=3/2,~I=1/2$ sector. To the right the two small peaks of the left figure magnified.}\label{fig:res31}%
\end{figure}

For the $J=3/2,~I=1/2$ sector, we show our results in Fig. \ref{fig:res31}.
From the results of $|T|^2$, we can also see three clear peaks around the range $4300 \sim 4500 \mev$, which are not far away below the thresholds of $\bar{D} \Sigma_c^*, ~\bar{D}^* \Sigma_c, ~\bar{D}^* \Sigma_c^*$ respectively. The strength of the second peak is 17 times bigger than the other two and the widths are small enough to allow the peaks to show up clearly. We have also checked that in these channels the real parts of the propagator $G$, Eq. \eqref{eq:G}, are acceptable too. So, these are our predictions for the new baryon states with total momentum $J=3/2$. We search the poles in the second Riemann sheet, and find $(4334.45+i19.41)\mev, ~(4417.04+i4.11)\mev, ~(4481.04+i17.38)\mev$. The couplings to each coupled channel corresponding to these poles are listed in Table \ref{tab:cou31}.
\begin{table}[ht]
     \renewcommand{\arraystretch}{1.2}
\centering
\caption{The coupling constants to various channels for certain poles in the $J=3/2,~I=1/2$ sector.} \label{tab:cou31}
\begin{tabular}{ccc ccc}
\hline\hline
$4334.45+i19.41$ & $J/\psi N$ & $\bar{D}^* \Lambda_c$ & $\bar{D}^* \Sigma_c$ & $\bar{D} \Sigma_c^*$ & $\bar{D}^* \Sigma_c^*$  \\
\hline
$g_i$ & $1.31-i0.18$ & $0.16-i0.23$ & $0.20-i0.48$ & $2.97-i0.36$ & $0.24-i0.76$  \\
$|g_i|$ & $1.32$ & $0.28$ & $0.52$ & $2.99$ & $0.80$  \\
\hline
$4417.04+i4.11$ & $J/\psi N$ & $\bar{D}^* \Lambda_c$ & $\bar{D}^* \Sigma_c$ & $\bar{D} \Sigma_c^*$ & $\bar{D}^* \Sigma_c^*$  \\
\hline
$g_i$ & $0.53-i0.07$ & $0.08-i0.07$ & $2.81-i0.07$ & $0.12-i0.10$ & $0.11-i0.51$  \\
$|g_i|$ & $0.53$ & $0.11$ & $2.81$ & $0.16$ & $0.52$  \\
\hline
$4481.04+i17.38$ & $J/\psi N$ & $\bar{D}^* \Lambda_c$ & $\bar{D}^* \Sigma_c$ & $\bar{D} \Sigma_c^*$ & $\bar{D}^* \Sigma_c^*$  \\
\hline
$g_i$ & $1.05+i0.10$ & $0.18-i0.09$ & $0.12-i0.10$ & $0.22-i0.05$ & $2.84-i0.34$  \\
$|g_i|$ & $1.05$ & $0.20$ & $0.16$ & $0.22$ & $2.86$  \\
\hline
\end{tabular}
\end{table}
From Table \ref{tab:cou31}, we find that the first pole, $(4334.45+i19.41)\mev$, couples most strongly to the channel $\bar{D} \Sigma_c^*$ and corresponds to a $\bar{D} \Sigma_c^*$ state, bound by $51\mev$ with respect to its threshold of $4385.3\mev$, decaying essentially into $J/\psi N$. The state corresponding to the big peak in Fig. \ref{fig:res31} (left) couples mostly to $\bar{D}^* \Sigma_c$, it is bound by $45\mev$ with respect to the threshold of this channel, $4462.2\mev$ and decays mostly into $J/\psi N$. The third state with $J=3/2,~I=1/2$ couples mostly to $\bar{D}^* \Sigma_c^*$, is bound by $45\mev$ with respect to the threshold of this channel, $4526.7\mev$ and also decays mostly into $J/\psi N$.

\begin{figure}[tb]
\epsfig{file=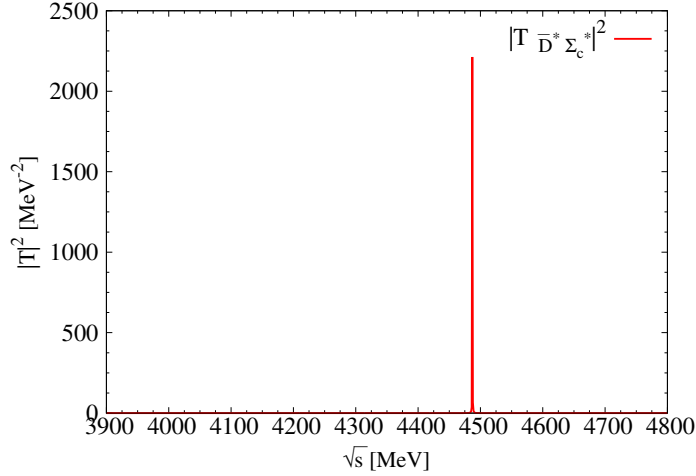, width=9cm}
\caption{The results of $|T|^2$ for the $J=5/2,~I=1/2$ sector.}\label{fig:res51}%
\end{figure}

Finally, we also find a new bound state of $\bar{D}^* \Sigma_c^*$ around $(4487.10+i0)\mev$ in the $J=5/2,~I=1/2$ sector, seen in Fig. \ref{fig:res51}.
As we can see in the figure, the state has no width, as it corresponds to a single channel, $\bar{D}^* \Sigma_c^*$ of Eq. \eqref{eq:ji51}. It is then a bound state in this channel. The pole appears in the first Riemann sheet and the state is bound by about $40\mev$ with respect to the $\bar{D}^* \Sigma_c^*$ threshold.

The states that we have reported are different states since they correspond to different energies or different total spin $J$. Hence, we get seven states. Yet, we found that some states of a given meson and baryon appear at about the same energy but different $J$. This was to be expected from the hidden gauge dynamics because for $\bar{D}^* B$ the main diagonal terms have an interaction of the type $\vec{\epsilon}~\vec{\epsilon}~'$, which is spin independent. Then, up to the small mixing with other channels, we get states degenerate in $J=1/2,~3/2$ for $\bar{D}^* \Sigma_c$ and $J=1/2,~3/2,~5/2$ for $\bar{D}^* \Sigma_c^*$. From this perspective we can present our results as saying that we get four bound states with about $40-50\mev$ binding, corresponding to $\bar{D} \Sigma_c,~\bar{D} \Sigma_c^*$ with $J=1/2,~3/2$ respectively, $\bar{D}^* \Sigma_c$ degenerated with $J=1/2,~3/2$ and $\bar{D}^* \Sigma_c^*$ degenerated with $J=1/2,~3/2,~5/2$.

The results reported in \cite{Garcia-Recio:2013gaa} show certain parallelism with those found here. There, seven odd parity $N-$like states are also found (three with spin 1/2 and 3/2 and a further one with spin 5/2). Moreover, the dynamics of these resonances is strongly influenced by the ${\bar D}^{(*)}\Sigma_c^{(*)}$ components, as it is the case here. Their masses, however, are quite, different, since those found in \cite{Garcia-Recio:2013gaa} lie in the region of 4 GeV, being thus significantly lighter than those found in this work. Besides differences of dynamical origin (SU(8) extension of the WT interaction + pattern of spin-flavor symmetry breaking versus SU(4) extension of the hidden gauge approach + pattern of flavor symmetry breaking) that can help to understand these changes in the position of the masses, there exists a major difference among both approaches in what concerns to the renormalization of the loop function, $G(s)$, in the coupled channels space. The baryon-meson propagator is logarithmically ultraviolet divergent, thus, the loop needed to be renormalized. Here, we use Eq.(64) with a scale $\mu=1000$ MeV and the subtraction constant $a(\mu)$ is set to $-2.3$, as it was done in \cite{wuprl,wuprc}. However in \cite{Garcia-Recio:2013gaa}, a subtraction point regularization is chosen such that $G_{ii} (s) = 0$ at a certain point $\sqrt{s} = \mu_{I}$. The subtraction point $\mu_{I}$ is set to $\sqrt{m^2_{\rm th} +M^2_{\rm th}}$, where $m_{\rm th}$ and $M_{\rm th}$ are, respectively, the masses of the meson and baryon producing the lowest threshold (minimal value of $m_{\rm th} +M_{\rm th}$) for each $I$ (isospin) sector, independent of the angular momentum $J$. This renormalization scheme was first proposed in Refs. \cite{Hofmann:2005sw,Hofmann:2006qx} and it was successfully used in Refs. \cite{GarciaRecio:2003ks,Gamermann:2011mq} for three light flavors and in the open charm (bottom) studies carried out in \cite{GarciaRecio:2008dp,Gamermann:2010zz,Romanets:2012hm} (\cite{GarciaRecio:2012db}). Both renormalization schemes (the one used in this work and that employed in \cite{Garcia-Recio:2013gaa}) lead to similar results for the case of light flavors, but however produce quite different results in the hidden-charm sector studied here. Indeed, a significant part of the differences between the masses of the resonances found here and those reported in \cite{Garcia-Recio:2013gaa} can be attributed to the different renormalization procedure followed in both works. As an example, let us pay attention to the $I=1/2$, $J=5/2$ sector, where there is only one coupled channel: ${\bar D}^* \Sigma^*_c$. The interactions used here and that of Ref. \cite{Garcia-Recio:2013gaa} are attractive and nearly the same, once it is taken into account that for this particular channel $1/f^2$ is replaced by $1/f^2_{D^*}$ in \cite{Garcia-Recio:2013gaa} according to pattern of spin-flavor symmetry breaking implemented in that work (see Sect. IIIB of \cite{Garcia-Recio:2013gaa}). However, the state found in \cite{Garcia-Recio:2013gaa} is around 450 MeV lighter (more bound) than that predicted here. This difference can be only attributed to the renormalization scheme\footnote{We should note that in \cite{Garcia-Recio:2013gaa}, the interaction is splitted in different irreducible representations of the symmetry group, and only those sectors where the interaction is attractive are studied.}. Large binding energies cannot be discarded. For instance, interpreting the $\Lambda_c (2595)$ in the open charm sector  as a ${\bar D}^* N$ bound state \cite{GarciaRecio:2008dp,Romanets:2012hm} would lead to a binding energy of around 350 MeV. On the other hand, the subtraction constant (main difference of the two renormalization schemes) generates terms at the next order of the expansion  used to determine the potential \cite{Nieves:1999bx}. Thus, different values for the subtraction constant can only be discriminated with the help of some phenomenological input, for instance the position of some states, that could be used to constrain such terms. 

We would like to finish this discussion just stressing again that, ignoring the difference in the mass positions, the isospin 1/2 states found in this work have a clear resemblance with those reported in \cite{Garcia-Recio:2013gaa}. The predicted new resonances definitely cannot be accommodated by quark models with three constituent quarks and they might be looked for in the forthcoming PANDA experiment at the future FAIR facility.

\section{Conclusions}

In this paper we have addressed a relevant topic which is to show the consistency of the dynamics of the local hidden gauge Lagrangians extrapolated to SU(4) with the LO constraints of Heavy Quark Spin Symmetry. These latter constraints are very powerful since they have their root in the QCD Lagrangian and must be understood as a very stringent. To show the consistency we have addressed the problem of the interaction of mesons and baryons with hidden charm, a problem of much current interest in view of ongoing work at the BES, BELLE, FAIR and other facilities. Once again the requirements of HQSS demanded that we put together pseudoscalar and vector mesons, as well as baryons with $J=1/2, 3/2$.  A series of relationships were developed for the transition potentials between the different meson-baryon channels in different combinations of spin and isospin. After this, we evaluated these matrix elements using the dynamics of the local hidden gauge approach and found them to fulfil all the relationships of LO HQSS, while at the same time providing some determined expressions for them which allowed us to find out the existence of several bound states or resonances stemming from this interaction. We found seven states with different energies or different spin-isospin quantum numbers. Yet, the fact that the interaction that we had for vector-baryon factorizes as $\epsilon \cdot \epsilon'$ produces matrix elements which are degenerate in the different spins allowed by the meson-baryon combinations. Hence, up to some different mixing with subleading channels, we found a very approximate degeneracy in the states that qualify as quasibound $\bar D^* B$. In view of this, the seven states that we found could be more easily classified as four basic states corresponding to a quasibound $\bar D \Sigma_c$ state which appears in $J=1/2$, a $\bar D \Sigma_c^*$ state in $J=3/2$, a  $\bar D^* \Sigma_c$ state which appears nearly degenerate in $J=1/2, ~3/2$ and a $\bar D^* \Sigma_c^*$ state which appears nearly degenerate in $J=1/2, ~3/2, ~5/2$. All the states are bound with about 50 MeV with respect to the corresponding $\bar D B$ thresholds and the width, except for the $J=5/2$ state, is also of the same order of magnitude. The $J=5/2$ state which appears in the single $\bar D^* \Sigma_c^*$ channel has the peculiarity that it has zero width in the space of states chosen. All the states found appear in $I=1/2$ and we found no states in $I=3/2$. The masses obtained here are substantially heavier than those found in another model \cite{Garcia-Recio:2013gaa} that also fulfils HQSS, but incorporates some elements of extra SU(8) symmetry and a different renormalization scheme. In spite of this, this latter model predicts also the same $J^P$ quantum numbers for these states. Experiments to search for the states predicted will bring further light on this issue in the future and we can only encourage them.

\section*{Acknowledgments}  
One of us, C. W. X., thanks C. Hanhart, F.K. Guo for helpful discussions in the Hadron Physics Summer School 2012 (HPSS2012), Germany.
This work is partly supported by the Spanish Ministerio de Economia y Competitividad and European FEDER funds under the contract number FIS2011-28853-C02-01 and FIS2011-28853-C02-02, and the Generalitat Valenciana in the program Prometeo, 2009/090. We acknowledge the support of the European Community-Research Infrastructure Integrating Activity Study of Strongly Interacting Matter (acronym Hadron Physics 3, Grant Agreement n. 283286) under the Seventh Framework Programme of EU.


\begin{thebibliography}{99} 

%\cite{Wu:2010jy}
\bibitem{wuprl} 
  J.~-J.~Wu, R.~Molina, E.~Oset and B.~S.~Zou,
  %``Prediction of narrow $N^*$ and $\Lambda^*$ resonances with hidden charm above 4 GeV,''
  Phys.\ Rev.\ Lett.\  {\bf 105}, 232001 (2010)
  [arXiv:1007.0573 [nucl-th]].
  %%CITATION = ARXIV:1007.0573;%%
  
  %\cite{Wu:2010vk}
\bibitem{wuprc} 
  J.~-J.~Wu, R.~Molina, E.~Oset and B.~S.~Zou,
  %``Dynamically generated $N^{*}$ and $\Lambda^*$ resonances in the hidden charm sector around 4.3 GeV,''
  Phys.\ Rev.\ C {\bf 84}, 015202 (2011)
  [arXiv:1011.2399 [nucl-th]].
  %%CITATION = ARXIV:1011.2399;%%
  
     %\cite{Bando:1984ej}
\bibitem{hidden1}
  M.~Bando, T.~Kugo, S.~Uehara, K.~Yamawaki and T.~Yanagida,
  %``Is Rho Meson A Dynamical Gauge Boson Of Hidden Local Symmetry?,''
  Phys.\ Rev.\ Lett.\  {\bf 54}, 1215 (1985).
  %%CITATION = PRLTA,54,1215;%%  
  
 
  
  %\cite{Bando:1987br}
\bibitem{hidden2}
  M.~Bando, T.~Kugo and K.~Yamawaki,
  %``Nonlinear Realization and Hidden Local Symmetries,''
  Phys.\ Rept.\  {\bf 164}, 217 (1988).
  %%CITATION = PRPLC,164,217;%%
  
 %\cite{Meissner:1987ge}
\bibitem{hidden4}
  U.~G.~Meissner,
  %``Low-Energy Hadron Physics From Effective Chiral Lagrangians With Vector
  %Mesons,''
  Phys.\ Rept.\  {\bf 161}, 213 (1988).
  %%CITATION = PRPLC,161,213;%%
 
%\cite{Weinberg:1978kz}
\bibitem{Weinberg:1978kz} 
  S.~Weinberg,
  %``Phenomenological Lagrangians,''
  Physica A {\bf 96}, 327 (1979).
  %%CITATION = PHYSA,A96,327;%%
  
  %\cite{Gasser:1983yg}
\bibitem{Gasser:1983yg} 
  J.~Gasser and H.~Leutwyler,
  %``Chiral Perturbation Theory to One Loop,''
  Annals Phys.\  {\bf 158}, 142 (1984).
  %%CITATION = APNYA,158,142;%%
  
  
 \bibitem{sakurai} J. J. Sakurai, Currents and Mesons (University of Chicago Press, Chicago, 1969). 
 
 %\cite{Ecker:1989yg}
\bibitem{Ecker:1989yg} 
  G.~Ecker, J.~Gasser, H.~Leutwyler, A.~Pich and E.~de Rafael,
  %``Chiral Lagrangians for Massive Spin 1 Fields,''
  Phys.\ Lett.\ B {\bf 223}, 425 (1989).
  %%CITATION = PHLTA,B223,425;%%
  
  
  %\cite{Molina:2008jw}
\bibitem{raquelrho} 
  R.~Molina, D.~Nicmorus and E.~Oset,
  %``The rho rho interaction in the hidden gauge formalism and the f(0)(1370) and f(2)(1270) resonances,''
  Phys.\ Rev.\ D {\bf 78}, 114018 (2008)
  [arXiv:0809.2233 [hep-ph]].
  %%CITATION = ARXIV:0809.2233;%%

%\cite{Geng:2008gx}
\bibitem{gengvec} 
  L.~S.~Geng and E.~Oset,
  %``Vector meson-vector meson interaction in a hidden gauge unitary approach,''
  Phys.\ Rev.\ D {\bf 79}, 074009 (2009)
  [arXiv:0812.1199 [hep-ph]].
  %%CITATION = ARXIV:0812.1199;%%
  
 %\cite{Nagahiro:2008um}
\bibitem{junko} 
  H.~Nagahiro, J.~Yamagata-Sekihara, E.~Oset, S.~Hirenzaki and R.~Molina,
  %``The gamma gamma decay of the f(0)(1370) and f(2)(1270) resonances in the hidden gauge formalism,''
  Phys.\ Rev.\ D {\bf 79}, 114023 (2009)
  [arXiv:0809.3717 [hep-ph]].
  %%CITATION = ARXIV:0809.3717;%% 
  
  
 %\cite{Branz:2009cv}
\bibitem{tanyageng} 
  T.~Branz, L.~S.~Geng and E.~Oset,
  %``Two-photon and one photon-one vector meson decay widths of the f(0)(1370), f(2)(1270), f(0)(1710), f-prime(2)(1525), and K*(2)(1430),''
  Phys.\ Rev.\ D {\bf 81}, 054037 (2010)
  [arXiv:0911.0206 [hep-ph]].
  %%CITATION = ARXIV:0911.0206;%% 
  
  
  %\cite{MartinezTorres:2009uk}
\bibitem{conchinos} 
  A.~Martinez Torres, L.~S.~Geng, L.~R.~Dai, B.~X.~Sun, E.~Oset and B.~S.~Zou,
  %``Study of the J/psi ---> phi(omega) f(2)(1270), J/psi ---> phi (omega) f-prime(2)(1525) and J/psi ---> K*0(892) anti-K*0(2) (1430) decays,''
  Phys.\ Lett.\ B {\bf 680}, 310 (2009)
  [arXiv:0906.2963 [nucl-th]].
  %%CITATION = ARXIV:0906.2963;%%
  
  %\cite{Geng:2009iw}
\bibitem{radiative} 
  L.~S.~Geng, F.~K.~Guo, C.~Hanhart, R.~Molina, E.~Oset and B.~S.~Zou,
  %``Study of the f(2)(1270), f(2)-prime(1525), f(0)(1370) and f(0)(1710) in the J/psi radiative decays,''
  Eur.\ Phys.\ J.\ A {\bf 44}, 305 (2010)
  [arXiv:0910.5192 [hep-ph]].
  %%CITATION = ARXIV:0910.5192;%%
  
  %\cite{Molina:2009eb}
\bibitem{raquelhideko} 
  R.~Molina, H.~Nagahiro, A.~Hosaka and E.~Oset,
  %``Scalar, axial-vector and tensor resonances from the rho D*, omega D* interaction in the hidden gauge formalism,''
  Phys.\ Rev.\ D {\bf 80}, 014025 (2009)
  [arXiv:0903.3823 [hep-ph]].
  %%CITATION = ARXIV:0903.3823;%%
  
  
  %\cite{Molina:2009ct}
\bibitem{xyz} 
  R.~Molina and E.~Oset,
  %``The Y(3940), Z(3930) and the X(4160) as dynamically generated resonances from the vector-vector interaction,''
  Phys.\ Rev.\ D {\bf 80}, 114013 (2009)
  [arXiv:0907.3043 [hep-ph]].
  %%CITATION = ARXIV:0907.3043;%%
  
  %\cite{Molina:2010tx}
\bibitem{raqueltanya} 
  R.~Molina, T.~Branz and E.~Oset,
  %``A new interpretation for the $D^*_{s2}(2573)$ and the prediction of novel exotic charmed mesons,''
  Phys.\ Rev.\ D {\bf 82}, 014010 (2010)
  [arXiv:1005.0335 [hep-ph]].
  %%CITATION = ARXIV:1005.0335;%%

%\cite{Dong:2012hc}
\bibitem{Dong:2012hc}
  Y.~Dong, A.~Faessler, T.~Gutsche and V.~E.~Lyubovitskij,
 %``Decays of Zb(+) and Zb'(+) as Hadronic Molecules,''
 J.\ Phys.\ G {\bf 40}, 015002 (2013)
 [arXiv:1203.1894 [hep-ph]].
 %%CITATION = ARXIV:1203.1894;%%
 %4 citations counted in INSPIRE as of 18 Mar 2013

%\cite{Branz:2010gd}
\bibitem{Branz:2010gd}
  T.~Branz, T.~Gutsche and V.~E.~Lyubovitskij,
 %``Two-photon decay of heavy hadron molecules,''
 Phys.\ Rev.\ D {\bf 82}, 054010 (2010)
 [arXiv:1007.4311 [hep-ph]].
 %%CITATION = ARXIV:1007.4311;%%
 %10 citations counted in INSPIRE as of 18 Mar 2013

%\cite{Branz:2010sh}
\bibitem{Branz:2010sh}
  T.~Branz, T.~Gutsche and V.~E.~Lyubovitskij,
 %``Hidden-charm and radiative decays of the Z(4430) as a hadronic D_1 \bar{D^\ast} bound state,''
 Phys.\ Rev.\ D {\bf 82}, 054025 (2010)
 [arXiv:1005.3168 [hep-ph]].
 %%CITATION = ARXIV:1005.3168;%%
 %10 citations counted in INSPIRE as of 18 Mar 2013

%\cite{Lee:2009hy}
\bibitem{Lee:2009hy}
  I.~W.~Lee, A.~Faessler, T.~Gutsche and V.~E.~Lyubovitskij,
 %``X(3872) as a molecular DD* state in a potential model,''
 Phys.\ Rev.\ D {\bf 80}, 094005 (2009)
 [arXiv:0910.1009 [hep-ph]].
 %%CITATION = ARXIV:0910.1009;%%
 %28 citations counted in INSPIRE as of 18 Mar 2013 
  
 %\cite{Sarkar:2009kx}
\bibitem{sourav} 
  S.~Sarkar, B.~-X.~Sun, E.~Oset and M.~J.~Vicente Vacas,
  %``Dynamically generated resonances from the vector octet-baryon decuplet interaction,''
  Eur.\ Phys.\ J.\ A {\bf 44}, 431 (2010)
  [arXiv:0902.3150 [hep-ph]].
  %%CITATION = ARXIV:0902.3150;%% 
  
  %\cite{Oset:2009vf}
\bibitem{angelsvec}
  E.~Oset, A.~Ramos,
  %``Dynamically generated resonances from the vector octet-baryon octet interaction,''
  Eur.\ Phys.\ J.\  {\bf A44}, 445-454 (2010).
  
  %\cite{Beringer:1900zz}
\bibitem{pdg} 
  J.~Beringer {\it et al.}  [Particle Data Group Collaboration],
  %``Review of Particle Physics (RPP),''
  Phys.\ Rev.\ D {\bf 86}, 010001 (2012).
  %%CITATION = PHRVA,D86,010001;%%

  %\cite{Garzon:2012np}
\bibitem{javier} 
  E.~J.~Garzon and E.~Oset,
  %``Effects of pseudoscalar-baryon channels in the dynamically generated vector-baryon resonances,''
  Eur.\ Phys.\ J.\ A {\bf 48}, 5 (2012)
  [arXiv:1201.3756 [hep-ph]].
  %%CITATION = ARXIV:1201.3756;%%
  

%\cite{Khemchandani:2011et}
\bibitem{kanchan1} 
  K.~P.~Khemchandani, H.~Kaneko, H.~Nagahiro and A.~Hosaka,
  %``Vector meson-Baryon dynamics and generation of resonances,''
  Phys.\ Rev.\ D {\bf 83}, 114041 (2011)
  [arXiv:1104.0307 [hep-ph]].
  %%CITATION = ARXIV:1104.0307;%%
  
%\cite{Khemchandani:2011mf}
\bibitem{kanchan2} 
  K.~P.~Khemchandani, A.~Martinez Torres, H.~Kaneko, H.~Nagahiro and A.~Hosaka,
  %``Coupling vector and pseudoscalar mesons to study baryon resonances,''
  Phys.\ Rev.\ D {\bf 84}, 094018 (2011)
  [arXiv:1107.0574 [nucl-th]].
  %%CITATION = ARXIV:1107.0574;%%
  
%\cite{Khemchandani:2012ur}
\bibitem{kanchan3} 
  K.~P.~Khemchandani, A.~Martinez Torres, H.~Nagahiro and A.~Hosaka,
  %``Negative parity $\Lambda$ and $\Sigma$ resonances coupled to pseudoscalar and vector mesons,''
  Phys.\ Rev.\ D {\bf 85}, 114020 (2012)
  [arXiv:1203.6711 [nucl-th]].
  %%CITATION = ARXIV:1203.6711;%%

   %\cite{Oset:2012ap}
\bibitem{review} 
  E.~Oset, A.~Ramos, E.~J.~Garzon, R.~Molina, L.~Tolos, C.~W.~Xiao, J.~J.~Wu and B.~S.~Zou,
  %``Interaction of vector mesons with baryons and nuclei,''
  Int.\ J.\ Mod.\ Phys.\ E {\bf 21}, 1230011 (2012)
  [arXiv:1210.3738 [nucl-th]].
  %%CITATION = ARXIV:1210.3738;%%
 


%\cite{Lutz:2003jw}
\bibitem{Lutz:2003jw} 
  M.~F.~M.~Lutz and E.~E.~Kolomeitsev,
  %``On charm baryon resonances and chiral symmetry,''
  Nucl.\ Phys.\ A {\bf 730}, 110 (2004)
  [hep-ph/0307233].
  %%CITATION = HEP-PH/0307233;%%
  
  

  
  %\cite{Lutz:2005ip}
\bibitem{Lutz:2005ip} 
  M.~F.~M.~Lutz and E.~E.~Kolomeitsev,
  %``Baryon resonances from chiral coupled-channel dynamics,''
  Nucl.\ Phys.\ A {\bf 755}, 29 (2005)
  [hep-ph/0501224].
  %%CITATION = HEP-PH/0501224;%%
  
    %\cite{Hofmann:2005sw}
\bibitem{Hofmann:2005sw} 
  J.~Hofmann and M.~F.~M.~Lutz,
  %``Coupled-channel study of crypto-exotic baryons with charm,''
  Nucl.\ Phys.\ A {\bf 763}, 90 (2005)
  [hep-ph/0507071].
  %%CITATION = HEP-PH/0507071;%%
  
   %\cite{Hofmann:2006qx}
\bibitem{Hofmann:2006qx} 
  J.~Hofmann and M.~F.~M.~Lutz,
  %``D-wave baryon resonances with charm from coupled-channel dynamics,''
  Nucl.\ Phys.\ A {\bf 776}, 17 (2006)
  [hep-ph/0601249].
  %%CITATION = HEP-PH/0601249;%%
  
  %\cite{Tolos:2004yg}
\bibitem{laura} 
  L.~Tolos, J.~Schaffner-Bielich and A.~Mishra,
  %``Properties of D-mesons in nuclear matter within a self-consistent coupled-channel approach,''
  Phys.\ Rev.\ C {\bf 70}, 025203 (2004)
  [nucl-th/0404064].
  %%CITATION = NUCL-TH/0404064;%%
  
  %\cite{Mizutani:2006vq}
\bibitem{Mizutani:2006vq} 
  T.~Mizutani and A.~Ramos,
  %``D mesons in nuclear matter: A DN coupled-channel equations approach,''
  Phys.\ Rev.\ C {\bf 74}, 065201 (2006)
  [hep-ph/0607257].
  %%CITATION = HEP-PH/0607257;%%
  
  %\cite{JimenezTejero:2009vq}
\bibitem{JimenezTejero:2009vq} 
  C.~E.~Jimenez-Tejero, A.~Ramos and I.~Vidana,
  %``Dynamically generated open charmed baryons beyond the zero range approximation,''
  Phys.\ Rev.\ C {\bf 80}, 055206 (2009)
  [arXiv:0907.5316 [hep-ph]].
  %%CITATION = ARXIV:0907.5316;%%

  
  %\cite{Haidenbauer:2007jq}
\bibitem{Haidenbauer:2007jq} 
  J.~Haidenbauer, G.~Krein, U.~-G.~Meissner and A.~Sibirtsev,
  %``Anti-D N interaction from meson-exchange and quark-gluon dynamics,''
  Eur.\ Phys.\ J.\ A {\bf 33}, 107 (2007)
  [arXiv:0704.3668 [nucl-th]].
  %%CITATION = ARXIV:0704.3668;%%
  
  %\cite{Haidenbauer:2008ff}
\bibitem{Haidenbauer:2008ff} 
  J.~Haidenbauer, G.~Krein, U.~-G.~Meissner and A.~Sibirtsev,
  %``Charmed meson rescattering in the reaction anti-p d ---> anti-D N,''
  Eur.\ Phys.\ J.\ A {\bf 37}, 55 (2008)
  [arXiv:0803.3752 [hep-ph]].
  %%CITATION = ARXIV:0803.3752;%%
  
  %\cite{Haidenbauer:2010ch}
\bibitem{Haidenbauer:2010ch} 
  J.~Haidenbauer, G.~Krein, U.~-G.~Meissner and L.~Tolos,
  %``DN interaction from meson exchange,''
  Eur.\ Phys.\ J.\ A {\bf 47}, 18 (2011)
  [arXiv:1008.3794 [nucl-th]].
  %%CITATION = ARXIV:1008.3794;%%
 
    
\bibitem{IW89}
  N.~Isgur and M.~B.~Wise,
  %``Weak Decays of Heavy Mesons in the Static Quark Approximation,''
  Phys.\ Lett.\ B {\bf 232}, 113 (1989).
  %%CITATION = PHLTA,B232,113;%%
  %1813 citations counted in INSPIRE as of 27 Feb 2013

\bibitem{Ne94}
  M.~Neubert,
  %``Heavy quark symmetry,''
  Phys.\ Rept.\  {\bf 245}, 259 (1994)
  [hep-ph/9306320].
  %%CITATION = HEP-PH/9306320;%%
  %1110 citations counted in INSPIRE as of 27 Feb 2013

\bibitem{MW00} A.V. Manohar and M.B. Wise, {\it Heavy Quark Physics},
  Cambridge Monographs on Particle Physics, Nuclear Physics and
  Cosmology, vol. 10
 

%\cite{GarciaRecio:2008dp}
\bibitem{GarciaRecio:2008dp} 
  C.~Garcia-Recio, V.~K.~Magas, T.~Mizutani, J.~Nieves, A.~Ramos, L.~L.~Salcedo and L.~Tolos,
  %``The s-wave charmed baryon resonances from a coupled-channel approach with heavy quark symmetry,''
  Phys.\ Rev.\ D {\bf 79}, 054004 (2009)
  [arXiv:0807.2969 [hep-ph]].
  %%CITATION = ARXIV:0807.2969;%%
  
  %\cite{Gamermann:2010zz}
\bibitem{Gamermann:2010zz} 
  D.~Gamermann, C.~Garcia-Recio, J.~Nieves, L.~L.~Salcedo and L.~Tolos,
  %``Exotic dynamically generated baryons with negative charm quantum number,''
  Phys.\ Rev.\ D {\bf 81}, 094016 (2010)
  [arXiv:1002.2763 [hep-ph]].
  %%CITATION = ARXIV:1002.2763;%%
  
  %\cite{Romanets:2012hm}
\bibitem{Romanets:2012hm} 
  O.~Romanets, L.~Tolos, C.~Garcia-Recio, J.~Nieves, L.~L.~Salcedo and R.~G.~E.~Timmermans,
  %``Charmed and strange baryon resonances with heavy-quark spin symmetry,''
  Phys.\ Rev.\ D {\bf 85}, 114032 (2012)
  [arXiv:1202.2239 [hep-ph]].
  %%CITATION = ARXIV:1202.2239;%%
  
%\cite{Garcia-Recio:2013gaa}
\bibitem{Garcia-Recio:2013gaa} 
  C.~Garcia-Recio, J.~Nieves, O.~Romanets, L.~L.~Salcedo and L.~Tolos,
  %``Hidden charm N and Delta resonances with heavy-quark symmetry,''
  arXiv:1302.6938 [hep-ph].
  %%CITATION = ARXIV:1302.6938;%%  


%\cite{GarciaRecio:2012db}
\bibitem{GarciaRecio:2012db} 
  C.~Garcia-Recio, J.~Nieves, O.~Romanets, L.~L.~Salcedo and L.~Tolos,
  %``Odd parity bottom-flavored baryon resonances,''
  Phys.\  Rev.\  D {\bf 87}, 034032 (2013)
  [arXiv:1210.4755 [hep-ph]].
  %%CITATION = ARXIV:1210.4755;%%
  %2 citations counted in INSPIRE as of 28 Feb 2013

%\cite{Aaij:2012da}
\bibitem{Aaij:2012da} 
  RAaij {\it et al.}  [LHCb Collaboration],
  %``Observation of excited Lambda_b0 baryons,''
  Phys.\ Rev.\ Lett.\  {\bf 109}, 172003 (2012)
  [arXiv:1205.3452 [hep-ex]].
  %%CITATION = ARXIV:1205.3452;%%
  %10 citations counted in INSPIRE as of 18 Apr 2013
  
   %\cite{Nieves:2012tt}
\bibitem{Nieves:2012tt} 
  J.~Nieves and M.~P.~Valderrama,
  %``The Heavy Quark Spin Symmetry Partners of the X(3872),''
  Phys.\ Rev.\ D {\bf 86}, 056004 (2012)
  [arXiv:1204.2790 [hep-ph]].
  %%CITATION = ARXIV:1204.2790;%%
  
    %\cite{HidalgoDuque:2012pq}
\bibitem{HidalgoDuque:2012pq} 
  C.~Hidalgo-Duque, J.~Nieves and M.~P.~Valderrama,
  %``Light Flavour and Heavy Quark Spin Symmetry in Heavy Meson Molecules,''
    Phys.\ Rev.\ D {\bf 87}, 076006 (2013)
  [arXiv:1210.5431 [hep-ph]].
  %%CITATION = ARXIV:1210.5431;%%
  
  %\cite{HidalgoDuque:2012ej}
\bibitem{HidalgoDuque:2012ej} 
  C.~Hidalgo-Duque, J.~Nieves and M.~P.~Valderrama,
  %``Heavy Quark Spin Symmetry and SU(3)-Flavour Partners of the X(3872),''
  arXiv:1211.7004 [hep-ph].
  %%CITATION = ARXIV:1211.7004;%%

%\cite{Guo:2009id}
\bibitem{Guo:2009id} 
  F.~-K.~Guo, C.~Hanhart and U.~-G.~Meissner,
  %``Implications of heavy quark spin symmetry on heavy meson hadronic molecules,''
  Phys.\ Rev.\ Lett.\  {\bf 102}, 242004 (2009)
  [arXiv:0904.3338 [hep-ph]].
  %%CITATION = ARXIV:0904.3338;%%

%\cite{Guo:2013sya}
\bibitem{Guo:2013sya} 
  F.~-K.~Guo, C.~Hidalgo-Duque, J.~Nieves and M.~P.~Valderrama,
  %``Consequences of Heavy Quark Symmetries for Hadronic Molecules,''
  arXiv:1303.6608 [hep-ph].
  %%CITATION = ARXIV:1303.6608;%%
  %5 citations counted in INSPIRE as of 18 Apr 2013


\bibitem{Rose}
  M.~E.~Rose, Elementary Theory of Angular Momentum, John Wiley, 1957.  


    %\cite{Nagahiro:2008cv}
\bibitem{hidekoroca}
  H.~Nagahiro, L.~Roca, A.~Hosaka and E.~Oset,
  %``Hidden gauge formalism for the radiative decays of axial-vector mesons,''
  Phys.\ Rev.\  D {\bf 79}, 014015 (2009)
  %%CITATION = PHRVA,D79,014015;%%
  
%\cite{Gamermann:2008jh}
\bibitem{Gamermann:2008jh} 
  D.~Gamermann, E.~Oset and B.~S.~Zou,
  %``The Radiative decay of psi(3770) into the predicted scalar state X(3700),''
  Eur.\ Phys.\ J.\ A {\bf 41}, 85 (2009)
  [arXiv:0805.0499 [hep-ph]].
  %%CITATION = ARXIV:0805.0499;%%  
  
 %\cite{Klingl:1997kf}
\bibitem{Klingl:1997kf} 
  F.~Klingl, N.~Kaiser and W.~Weise,
  %``Current correlation functions, QCD sum rules and vector mesons in baryonic matter,''
  Nucl.\ Phys.\ A {\bf 624}, 527 (1997)
  [hep-ph/9704398].
  %%CITATION = HEP-PH/9704398;%%
 

  %\cite{Palomar:2002hk}
\bibitem{Palomar:2002hk}
  J.~E.~Palomar and E.~Oset,
  %``The Phi N N coupling from chiral loops,''
  Nucl.\ Phys.\  A {\bf 716}, 169 (2003)
  %%CITATION = NUPHA,A716,169;%%  
  
  
  \bibitem{Eck95} G. Ecker, Prog. Part. Nucl. Phys. 35 (1995) 1.
 
\bibitem{Be95} V. Bernard, N. Kaiser and U. G. Meissner, Int. J. Mod. Phys.
E4 (1995) 193. 

  
  %\cite{Jenkins:1991es}
\bibitem{manohar}
  E.~E.~Jenkins and A.~V.~Manohar,
  %``Chiral corrections to the baryon axial currents,''
  Phys.\ Lett.\  B {\bf 259}, 353 (1991).
  %%CITATION = PHLTA,B259,353;%%

\bibitem{close}
  F.~E.~Close, An Introduction to Quarks and Partons, Academic Press, 1979.
  
    %\cite{Oset:1997it}
\bibitem{angels}
  E.~Oset and A.~Ramos,
  %``Non perturbative chiral approach to s-wave anti-K N interactions,''
  Nucl.\ Phys.\  A {\bf 635} (1998) 99.
 %%CITATION = NUPHA,A635,99;%% 

  
  %\cite{Oller:2000fj}
\bibitem{ollerulf}
 J.~A.~Oller and U.~G.~Meissner,
 %``Chiral dynamics in the presence of bound states: Kaon nucleon  interactions
 %revisited,''
 Phys.\ Lett.\  B {\bf 500}, 263 (2001).
 %%CITATION = PHLTA,B500,263;%%
  
 %\cite{Nieves:1999bx}
\bibitem{Nieves:1999bx} 
  J.~Nieves and E.~Ruiz Arriola,
  %``Bethe-Salpeter approach for unitarized chiral perturbation theory,''
  Nucl.\ Phys.\ A {\bf 679}, 57 (2000)
  [hep-ph/9907469].
  %%CITATION = HEP-PH/9907469;%%
  %97 citations counted in INSPIRE as of 14 Mar 2013
 
  
%\cite{Borasoy:1998pe}
\bibitem{Borasoy:1998pe} 
  B.~Borasoy,
  %``Baryon axial currents,''
  Phys.\ Rev.\ D {\bf 59}, 054021 (1999)
  [hep-ph/9811411].
  %%CITATION = HEP-PH/9811411;%%


%\cite{Gamermann:2009uq}
\bibitem{gamer} 
  D.~Gamermann, J.~Nieves, E.~Oset and E.~Ruiz Arriola,
  %``Couplings in coupled channels versus wave functions: application to the X(3872) resonance,''
  Phys.\ Rev.\ D {\bf 81}, 014029 (2010)
  [arXiv:0911.4407 [hep-ph]].
  %%CITATION = ARXIV:0911.4407;%%
  
%\cite{Wu:2010rv}
\bibitem{wuzou} 
  J.~-J.~Wu and B.~S.~Zou,
  %``Prediction of super-heavy $N^*$ and $\Lambda^*$ resonances with hidden beauty,''
  Phys.\ Lett.\ B {\bf 709}, 70 (2012)
  [arXiv:1011.5743 [hep-ph]].
  %%CITATION = ARXIV:1011.5743;%%
  
%\cite{Guo:2005wp}
\bibitem{Guo:2005wp} 
  F.~-K.~Guo, R.~-G.~Ping, P.~-N.~Shen, H.~-C.~Chiang and B.~-S.~Zou,
  %``S wave K pi scattering and effects of kappa in J/psi ---> anti-K*0 (892) K+ pi-,''
  Nucl.\ Phys.\ A {\bf 773}, 78 (2006)
  [hep-ph/0509050].
  %%CITATION = HEP-PH/0509050;%%  

%\cite{Roca:2005nm}
\bibitem{luis} 
  L.~Roca, E.~Oset and J.~Singh,
  %``Low lying axial-vector mesons as dynamically generated resonances,''
  Phys.\ Rev.\ D {\bf 72}, 014002 (2005)
  [hep-ph/0503273].
  %%CITATION = HEP-PH/0503273;%%
  
%\cite{YamagataSekihara:2010pj}
\bibitem{junko2}
  J.~Yamagata-Sekihara, J.~Nieves and E.~Oset,
 %``Couplings in coupled channels versus wave functions in the case of resonances: application to the two $\Lambda(1405)$ states,''
 Phys.\ Rev.\ D {\bf 83}, 014003 (2011)
 [arXiv:1007.3923 [hep-ph]].
 %%CITATION = ARXIV:1007.3923;%% 

%\cite{GarciaRecio:2003ks}
\bibitem{GarciaRecio:2003ks} 
  C.~Garcia-Recio, M.~F.~M.~Lutz and J.~Nieves,
  %``Quark mass dependence of s wave baryon resonances,''
  Phys.\ Lett.\ B {\bf 582}, 49 (2004)
  [nucl-th/0305100].
  %%CITATION = NUCL-TH/0305100;%%
  %136 citations counted in INSPIRE as of 18 Apr 2013

%\cite{Gamermann:2011mq}
\bibitem{Gamermann:2011mq} 
  D.~Gamermann, C.~Garcia-Recio, J.~Nieves and L.~L.~Salcedo,
  %``Odd Parity Light Baryon Resonances,''
  Phys.\ Rev.\ D {\bf 84}, 056017 (2011)
  [arXiv:1104.2737 [hep-ph]].
  %%CITATION = ARXIV:1104.2737;%%
  %12 citations counted in INSPIRE as of 19 Apr 2013
  
 
  
\end{thebibliography}
\end{document}